\documentclass[twoside,9pt,journal]{IEEEtran}
\ifCLASSINFOpdf
\else
    \usepackage[dvips]{graphicx}
\fi
\usepackage[T1]{fontenc}
\usepackage{textcomp}
\usepackage{gensymb}

\usepackage{subfigure}

\usepackage{amsmath,amssymb,amsfonts}
\usepackage{algorithmic}
\ifCLASSOPTIONcompsoc
\usepackage[caption=false,font=normalsize,labelfon
t=sf,textfont=sf]{subfig}
\else
\usepackage[caption=false,font=footnotesize]{subfi
g}

\usepackage[nocompress]{cite}
\ifCLASSINFOpdf
\else
\fi
\usepackage{booktabs}
\usepackage{multirow}
\usepackage{color}
\usepackage{array}

\usepackage{amsmath}

\usepackage{wasysym}

\usepackage{csquotes}

\usepackage{pgfplots}
\usepackage[RPvoltages]{circuitikz}
\usepackage{upgreek}
\pgfplotsset{compat=1.17} 

\begin{document}

\bstctlcite{modify_bib}


%\title{Dual-wideband Dual-polarized Magnetoelectric Dipole Antenna with Second-order Bandstop Filtering Characteristic for 5G Millimeter-Wave Applications}
%\title{Integration of Second-order Bandstop Filter into Dual-polarized 5G Millimeter-Wave Antenna}
%\title{A Broadband Cross-Slotted Patch Antenna With Dual Polarization for Millimeter-Wave Base Station Applications}

\title{%{\fontsize{24}{26}\selectfont{Communication\rule{29.9pc}{0.5pt}}}\break\fontsize{16}{18}\selectfont
Integration of Second-Order Bandstop Filter Into a Dual-Polarized 5G Millimeter-Wave Magneto-Electric Dipole Antenna}


\author{%Jiangcheng~Chen,~\IEEEmembership{Graduate Student Member,~IEEE,}
        %Markus~Berg,
        %Kimmo~Rasilainen,~\IEEEmembership{Member,~IEEE,}
        %Zeeshan~Siddiqui,~\IEEEmembership{Graduate Student Member,~IEEE,}
        %Marko~E.~Leinonen,~\IEEEmembership{Member,~IEEE,}\\ and~Aarno~Pärssinen,~\IEEEmembership{Senior Member,~IEEE}% <-this % stops a space

Jiangcheng~Chen, Markus~Berg, Kimmo~Rasilainen, Zeeshan~Siddiqui, Marko~E.~Leinonen, and Aarno~Pärssinen% <-this % stops a space

\thanks{Manuscript received 30 October 2023. %; revised Day Month 202X; accepted Day Month 202X. Date of publication Day Month 202X; date of current version Day Month 202X.  
%.; accepted Month Day, Year. Date of publication Month Day, Year; date of current version Month Day, Year.} 
This work was supported in part by Nokia Corporation Ltd., in part by Research Council of Finland through the 6G Flagship program under Grant 346208 and in part by Business Finland RF Sampo project under Grant 2993/31/2021. The work of Jiangcheng Chen was supported in part by the Nokia Foundation, by the HPY Research Foundation, and by the Riitta and Jorma J. Takanen Foundation. (\emph{Corresponding author: Jiangcheng Chen}.)}% 
\thanks{J. Chen, K. Rasilainen, Z. Siddiqui, M. E. Leinonen, and A. P{\"a}rssinen are with the Centre for Wireless Communications, University of Oulu, FI-90570 Oulu, Finland. M. Berg is currently with ExcellAnt Ltd., FI-90590 Oulu, Finland.
 (e-mail: jiangcheng.chen@oulu.fi).}% <-this % stops a space
%\thanks{Color versions of one or more figures in this communication are available at https://doi.org/10.1109/TAP.XXXX.XXXXXXX.}
%\thanks{Digital Object Identifier 10.1109/TAP.XXXX.XXXXXXX}
}

% \markboth{IEEE Transactions on Antennas and Propagation}%
% {Chen \MakeLowercase{\textit{et al.}}: Integration of Second-Order Bandstop Filter Into a Dual-Polarized 5G MM-Wave ME Dipole Antenna}

\maketitle

\begin{abstract} This communication proposes a dual-wideband differentially fed dual-polarized magnetoelectric (ME) dipole with second-order bandstop filtering for millimeter-wave (mm-Wave) applications at 24.25--29.5\,GHz and 37--43.5\,GHz. Without disturbing the complementary antenna operation, two resonator types (hairpin and coupled ${\lambda}$/4 open-/short-circuited stub resonators), are embedded into the wideband ME dipole to create two transmission poles and two zeros for sharp band-edge selectivity. This allows independent manipulation of the transmission poles and zeros and a compact ME dipole size. Across the operating band, the symmetric filtering antenna design %\red{with differential feeding} 
has more than 31.6\,dB of port-to-port isolation. Measured results show symmetrical E- and H-plane radiation patterns and cross-polarization levels lower than --25.1\,dB. The measured gains of the single element and a 2\(\times\)2 array are 8.3\,dBi and 12.5\,dBi, respectively. Also, the band rejection reaches 23.7\,dB and 21.8\,dB for single element and array, respectively.
%\red{Something should be written here.}
%This article proposes a wideband differentially-fed dual-polarized magnetoelectric (ME) dipole for millimeter-wave (mm-Wave) applications. Various electric and magnetic characteristic modes of a slotted patch antenna are investigated and utilized effectively to create a stable broadside radiation pattern, covering 5G frequency bands from 24.25\,GHz to 40\,GHz. To implement this, the lifted ground (LGND) concept is applied to achieve a 57.1\% impedance bandwidth for a single antenna element. Additionally, the three resonances of the antenna can be manipulated independently. The use of differential feeding allows more than 36\,dB of port-to-port isolation across the entire operating band. The measured gains of the single element and 2\(\times\)2 array are 8.4\,dBi and 13.4\,dBi, respectively. Also, the measured results indicate symmetrical E- and H-plane radiation patterns and cross-polarization levels lower than --26\,dB. With the favorable electrical performance, compact size, simple structure and low-cost fabrication, the proposed ME dipole is a promising candidate for mm-Wave Antenna-in-Package (AiP) applications. 
\end{abstract} 
% is it necessary to mention two electric modes and one magnetric mode are selected for the application

% \fi
\IEEEpeerreviewmaketitle

\section{Introduction}

% {T}{his} demo file is intended to serve as a ``starter file''
% for IEEE journal papers produced under \LaTeX\ using

\IEEEPARstart{F}{ifth generation} (5G) wireless communications uses millimeter-wave (mm-Wave) frequencies for  %offer unprecedented spectrum to accessed mobile terminals, enabling 
high data rates, large channel capacity, and low latency~\cite{Rappaport5G2017}. %, Rappaport5G2013, Andrews5G2014}. 
Several mm-Wave bands at 24.25--29.5\,GHz and 37--43.5\,GHz have been allocated for 5G systems as 5G New Radio FR2 bands n257--n261 \cite{Yang2022}. Due to wide spectrum, broadband, dual-band or multiband mm-Wave antennas~\cite{Li2018,ZhangOverview2019,Kibaroglu2018} are needed to cover all desired subbands with a simple feed solution for miniaturization, ease of integration with Antenna-in-Package (AiP) technology, and low fabrication cost. Dual-polarized operation with high cross-polarization discrimination (XPD) and port-to-port isolation is also needed to increase channel capacity and reduce multipath fading effects using polarization diversity~\cite{Xue2013ME, Lian2016Xslot}.
%Generally, simultaneous operation with a compactness and broadband operation in each of the desired bands are the two fundamental technical challenges. 

%Recently, various mm-Wave antenna designs have been developed to meet the requirements. 
One solution is to cover the 5G FR2 lower-band (LB) subbands at 24.25--29.5\,GHz (19.5\%) and upper band (UB) at 37--43.5\,GHz (16.2\%) by a broadband antenna operating from 24.25 to 43.5\,GHz (56.8\%). For broader bandwidths (BW), one approach is to combine multimodes (e.g. higher order modes) by modifying the antenna structure while the lower modes remain stable, such as the circular patch \cite{zhang2020} and E-shaped patch \cite{Yin2019Epatch}. Another way is to stack multiple broadside radiators with a compact footprint \cite{Dzagbletey2018stacked}. 
These designs have insufficient bandwidth, and they are only single-polarized.% limited to only a single polarization.

Alternatively, the 5G FR2 LB and UB can be covered using a dual-wideband antenna. At mm-Wave frequencies, quarter-mode substrate-integrated waveguide (SIW) antennas \cite{Hong2019SIW, Deckmyn2019SIW, Sun2020SIW} have been used for dual-band operation with compact dimensions, low profile and ease of integration. However, owing to single-SIW mode operation, they can only cover the single subband i.e. 28-GHz (27.5–29.5\,GHz) and 38-GHz (37–38.6\,GHz) subbands at LB and UB, respectively. 
Another popular dual-band antenna is shared aperture antenna, such as an SIW slot array antenna \cite{Zhang2019shared}, a microstrip grid antenna with parasitic patches \cite{Xu2021shared},  a differentially fed slot antenna loaded with a dielectric resonator antenna \cite{Sun2016shared}, and a Combined-Ridge-Groove-Gap-Waveguide-fed shared circular aperture antenna \cite{Ferrando2019shared}. Still, these works cannot cover the required LB and UB simultaneously, with maximum bandwidth of 15.7\% and 16.7\% at their lower and upper operating bands, respectively.
Other dual-band antenna types (e.g., metasurface-based dual-band antennas \cite{Li2018,LiMeta2020, FengMeta2020}, gridded patch \cite{Sun2021Dualband}, scalable antenna \cite{Hu2020,Zeeshan2023}) show sufficient bandwidth, but for only one band at a time (either LB or UB). %, they cannot cover the other band. 
%\cite{Yang2022} 

%At mm-Wave frequencies, it is preferable to introduce filtering characteristics to reject the undesired frequency bands. 
Filtering antennas can be implemented to replace the filter in RF frontend, which reduces the insertion loss (see, e.g., \cite{Zhang2022}). 
%\cite{Chu2015, Yang2020, Zhang2022}. 
In \cite{Yang2022}, a U-shaped dipole with parasitic patch was proposed with bandstop filtering characteristic can fully cover the 5G NR mm-wave bands of n257--n261. However, the XPD and isolation are both at a level of 20\,dB, and the band-edge selectivity at the stopband is poor. The antenna in \cite{Zeeshan2023} also shows a bandstop response at the mid-band of LB and UB, but its rejection level is around 10\,dB.
 % is mid-band correct? shall we use in between
This communication presents a dual-polarized differential probe-fed ME dipole for 5G mm-Wave bands with simultaneous LB and UB operation. The proposed antenna combines a second-order bandstop filter with a recently reported complementary antenna \cite{Chen2022} containing two electric dipole modes and one magnetic dipole mode. The antenna has high band-edge selectivity and controllable bandwidth at stopband, and its main novelties can be summarized as follows: 
%\begin{enumerate}

1) Two physically isolated resonators (hairpin and coupled stubs with different ground references) allow independent control of the location of the transmission zeros/poles, stopband edge selectivity, band-rejection level, and bandwidth of the stopband.

2) The coupled stub uses broadside coupling for structural symmetry. With high quality factor 
$(Q)$ due to coupling between the open- and short-circuited stubs, it is possible to design a narrowband filter.

% 3) The design parameters of the hairpin and stub resonators can be altered to modify the bandwidth of the stopband. \red{can we merge this item to 1) since they are talking similar thing} \blue{Yes, I think so.}% to achieve various \
% %stopband bandwidth

3) By incorporating non-redundant transmission lines connecting the upper layer hairpin resonators and lower layer stub resonators into the filter design, the bandstop filter becomes quasi-optimal with steeper attenuation. 
    
4) Adding the two resonator sets improves the matching while not breaking the ME dipole antenna operation nor increasing its footprint. %\red{in ref, the bandstop ant has no additional resonance/matching improvement, but our design has one more matching point. The traditional way will increase the footprint}
%\end{enumerate}

%. In Section II, antenna designs of single element, and a 2\(\times\)2 array for verification of mutual coupling are presented. The operating principle and filter synthesis process for the two-pole optimum bandstop filter design is clearly revealed in Section III. It provides insights into the way to simultaneously improve the impedance matching and produce an additional transmission zero by adding a second coupled resonator. The simulated and measured results as well as parametric studies based on the equivalent circuit of the proposed band-notched ME dipole are demonstrated in Section IV, followed by conclusions in Section V.

\begin{figure}[!t]
%\centering
\hfill
\subfigure[]{\includegraphics[width=1.7in]{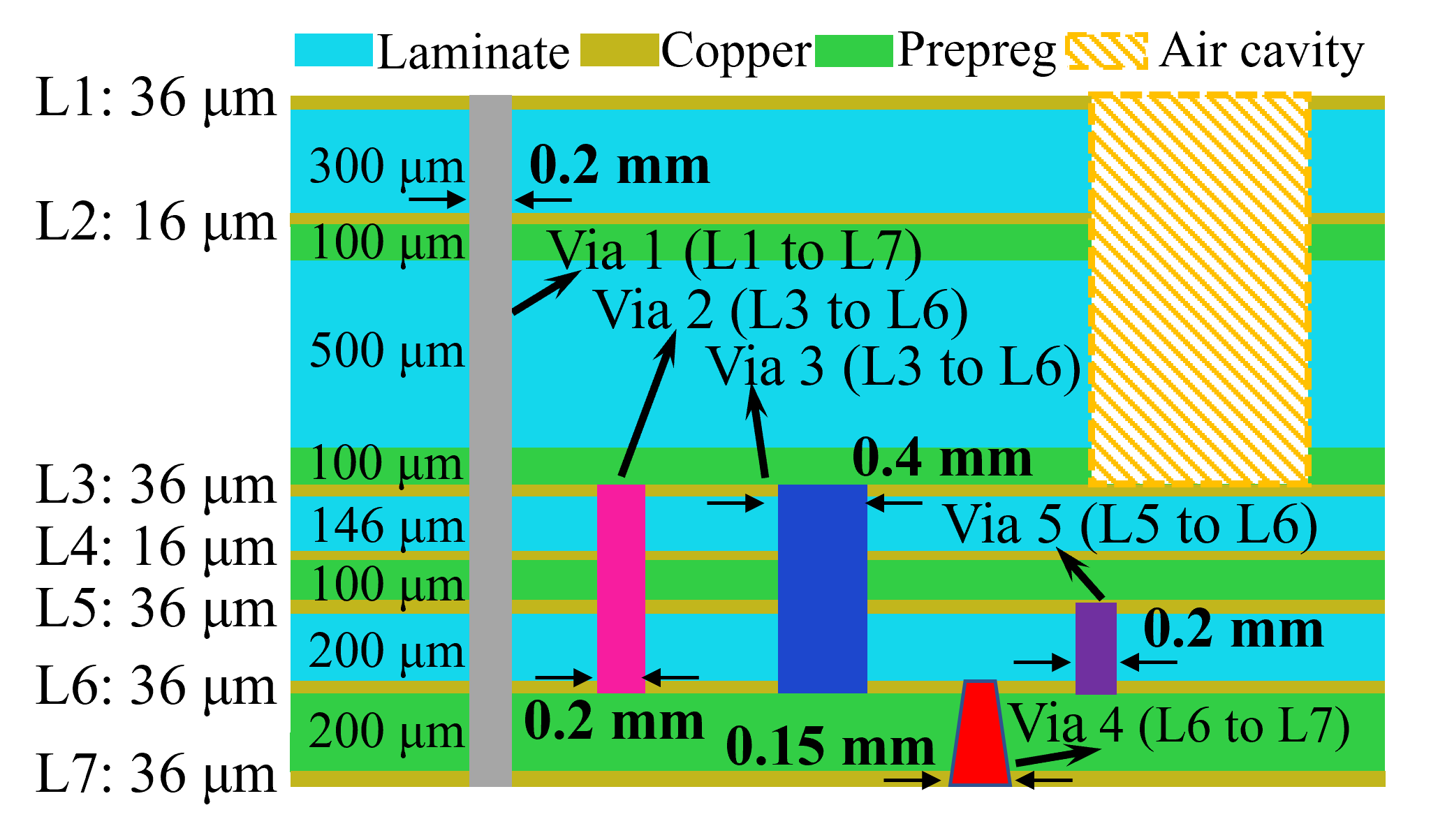}}
\hfill
%\centering
\subfigure[]{\includegraphics[scale=0.21]{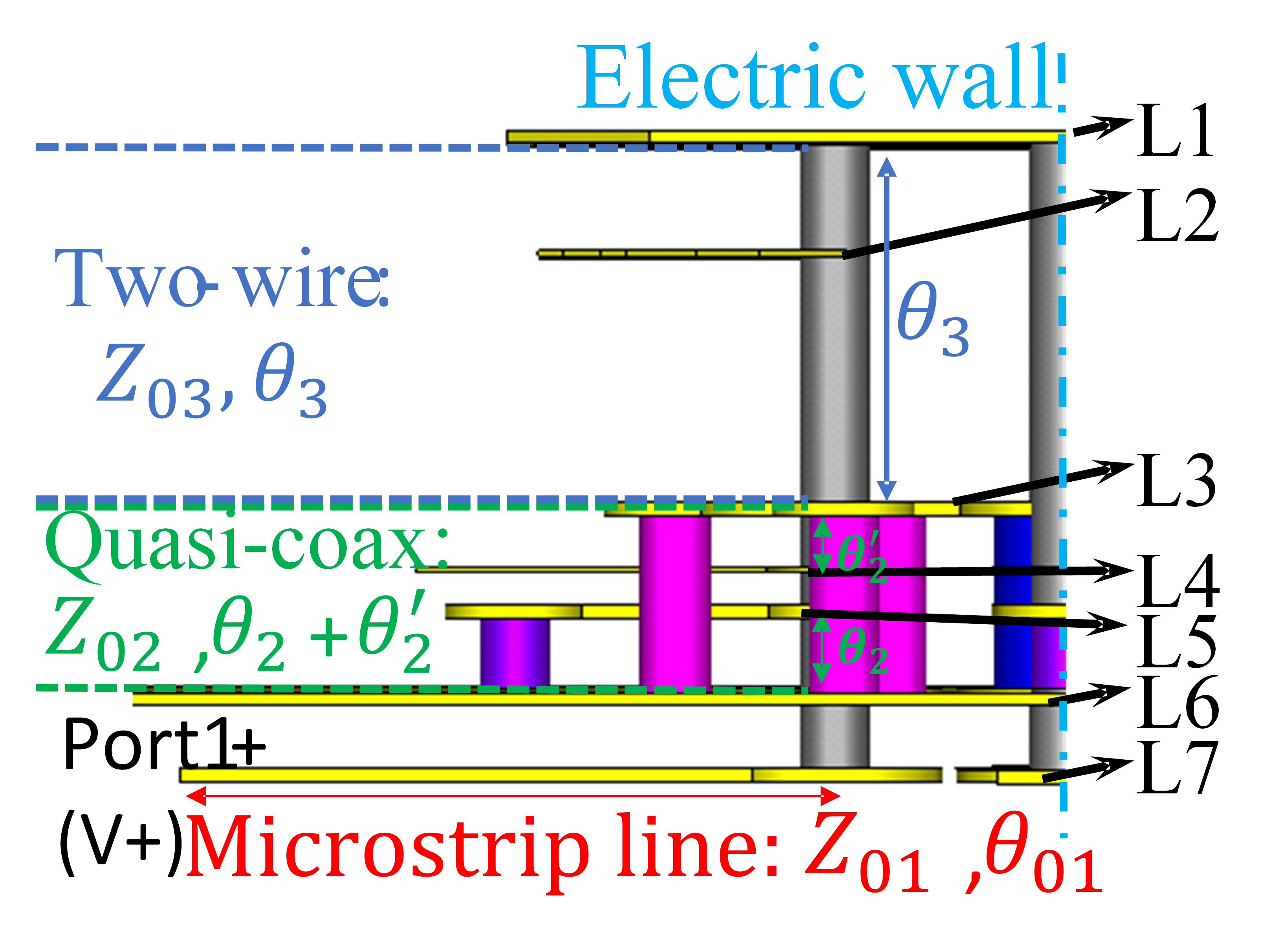}
}
\hfill
\subfigure[]{\includegraphics[scale=0.21]{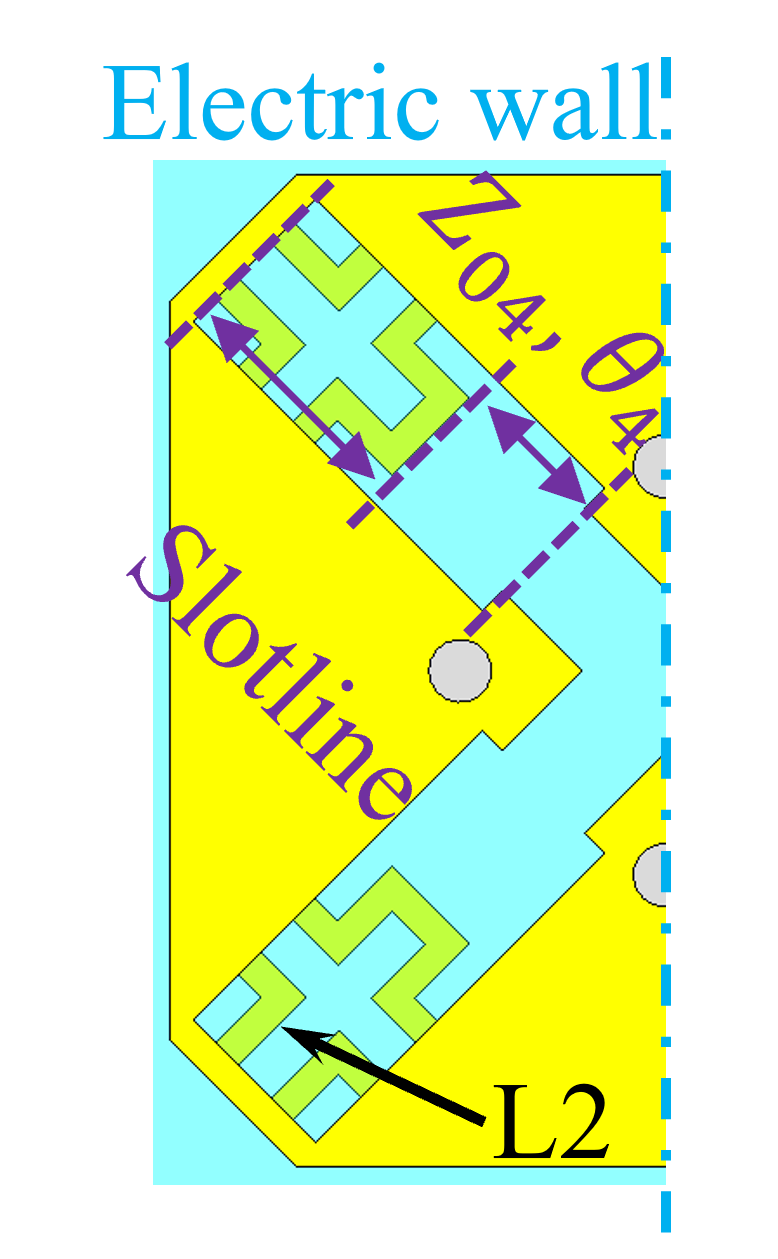}
}
\subfigure[Layer 1 (L1)]{\includegraphics[width=1.25in]{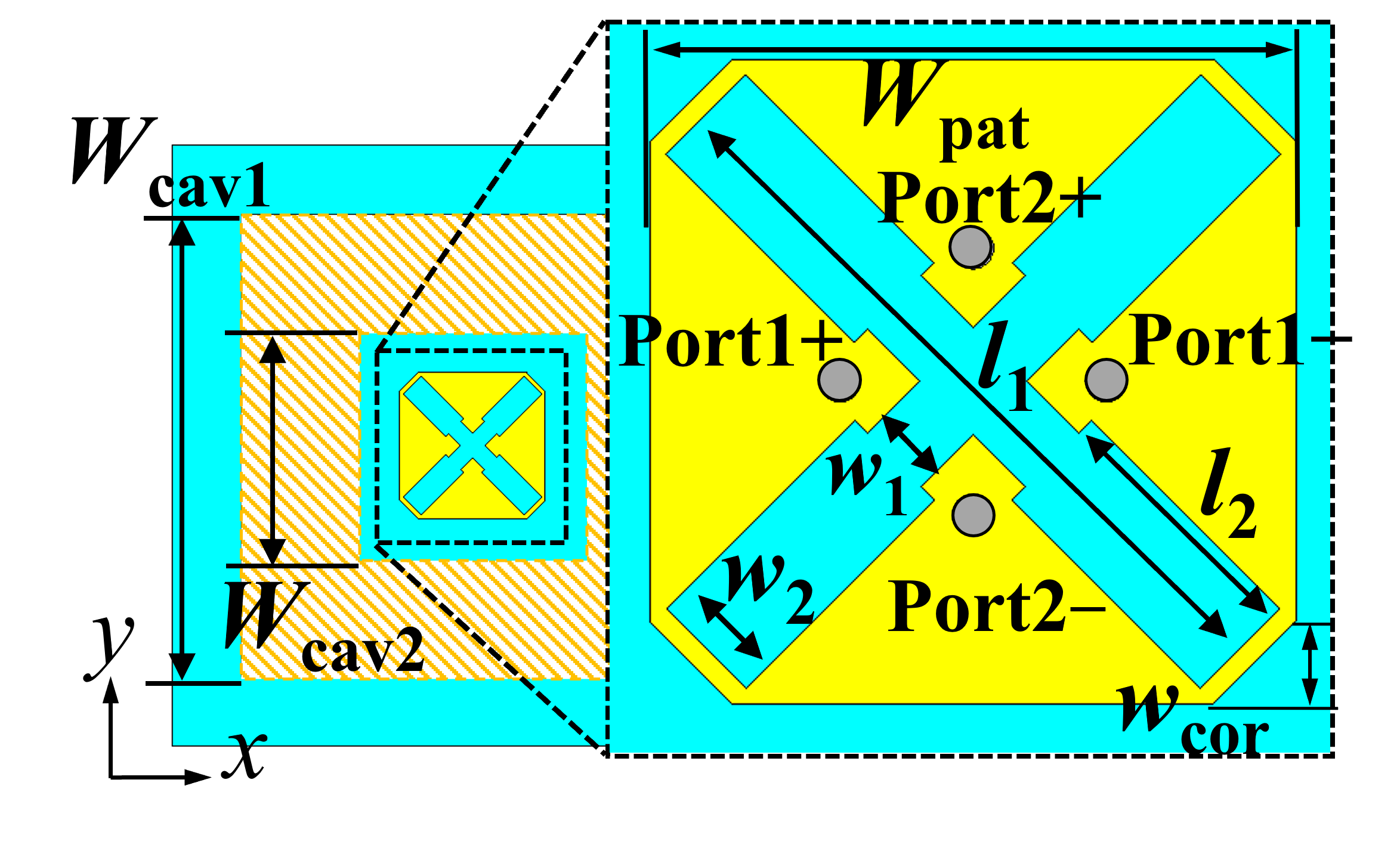}}
\hfill
\subfigure[Layer 2 (L2)]{\includegraphics[width=1.15in]{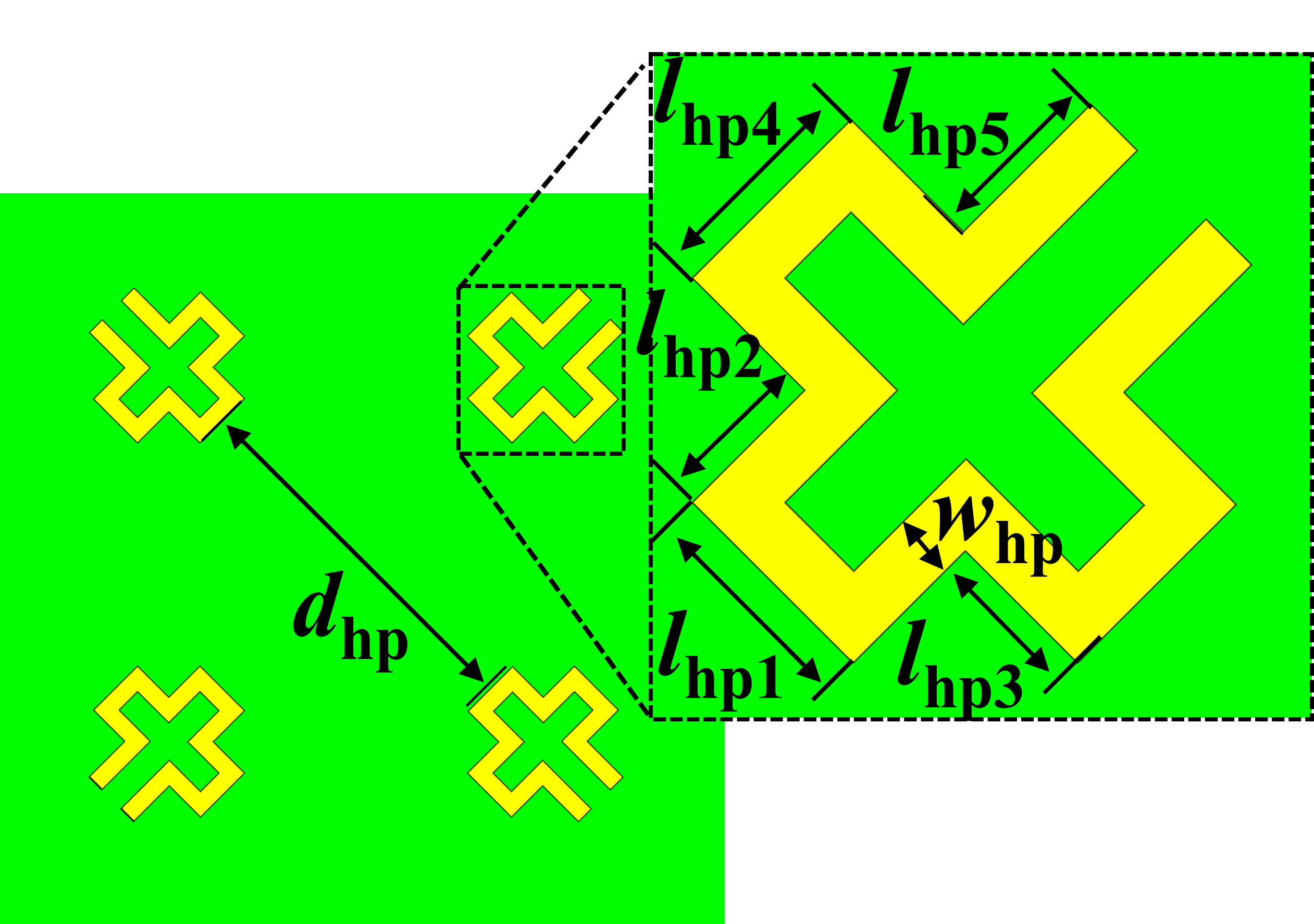}}
\hfill
\subfigure[Layer 3 (L3)]{\includegraphics[width=0.87in]{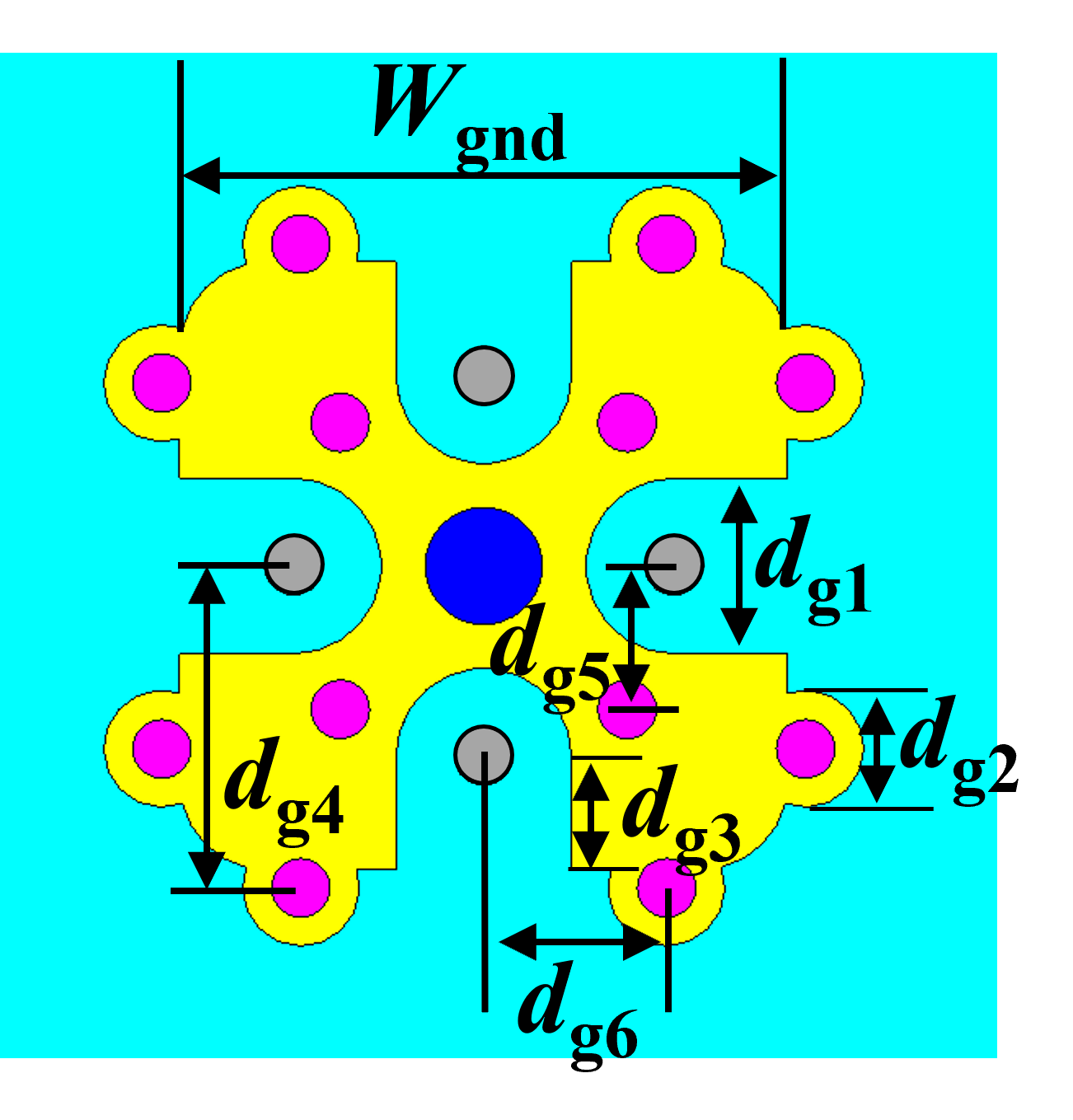}}
\hfill
\subfigure[Layer 4 (L4)]{\includegraphics[width=0.8in]{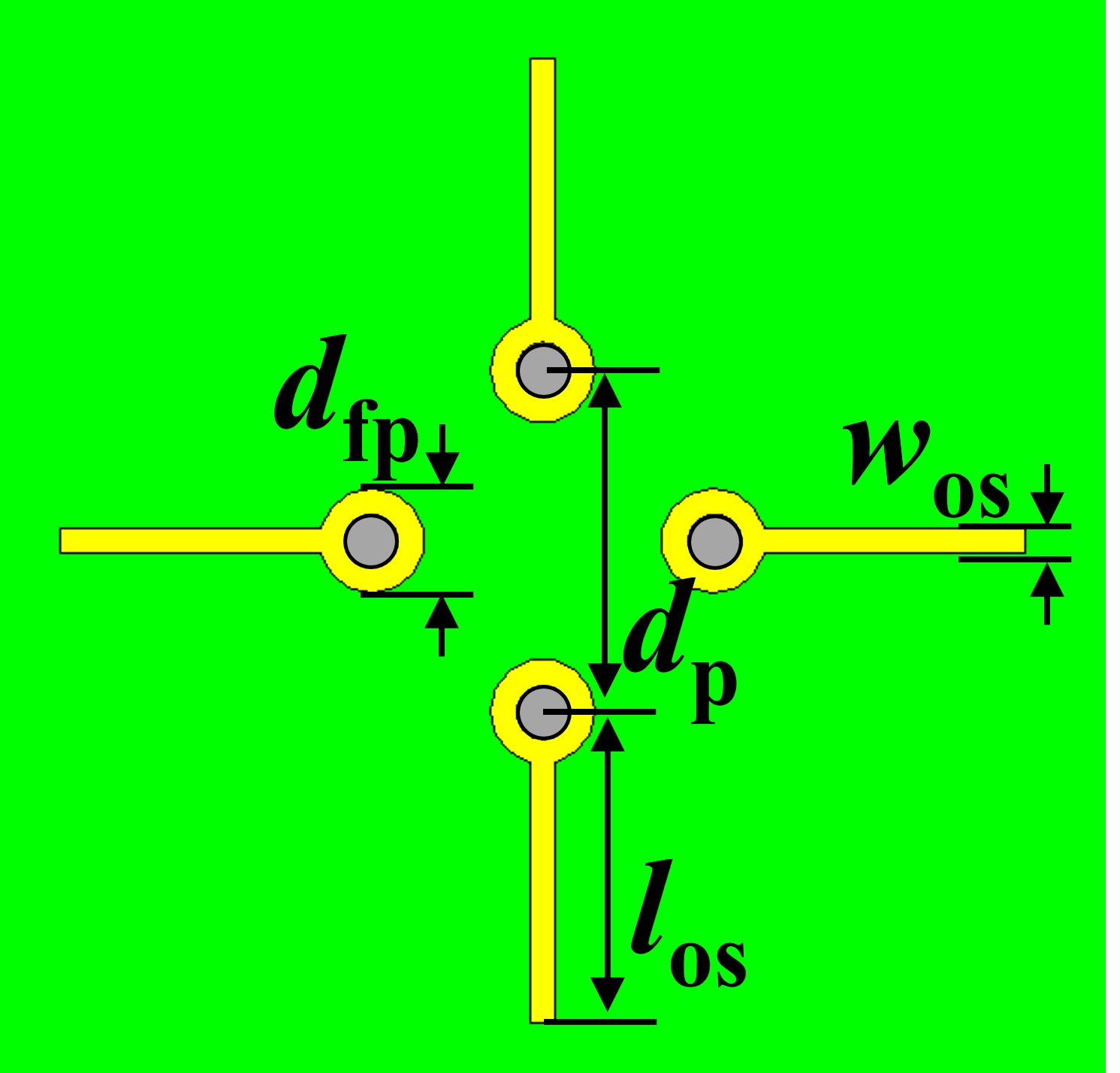}}
\hfill
\subfigure[Layer 5 (L5)]{\includegraphics[width=0.82in]{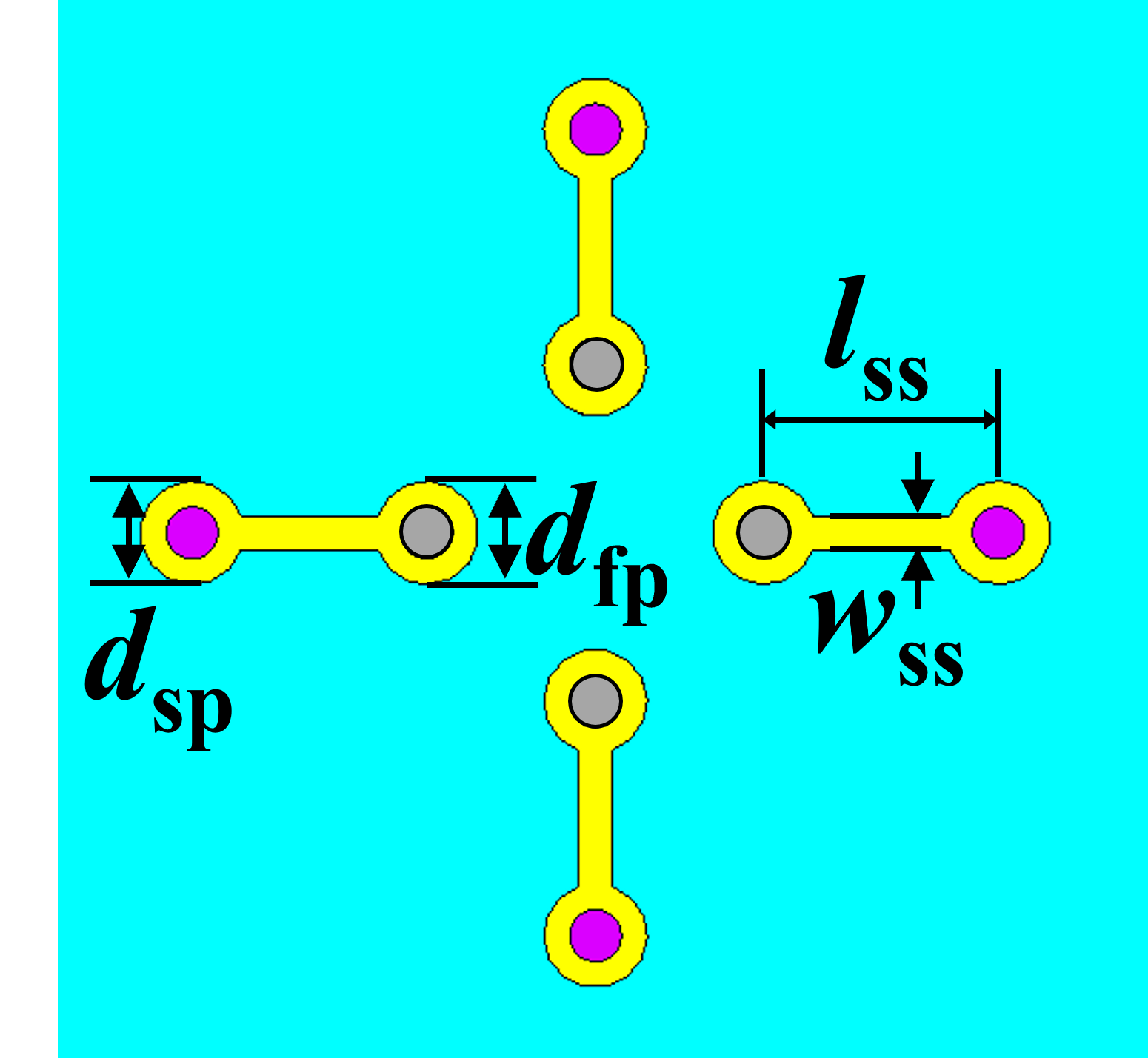}}
\hfill
\subfigure[Layer 6 (L6)]{\includegraphics[width=0.8in]{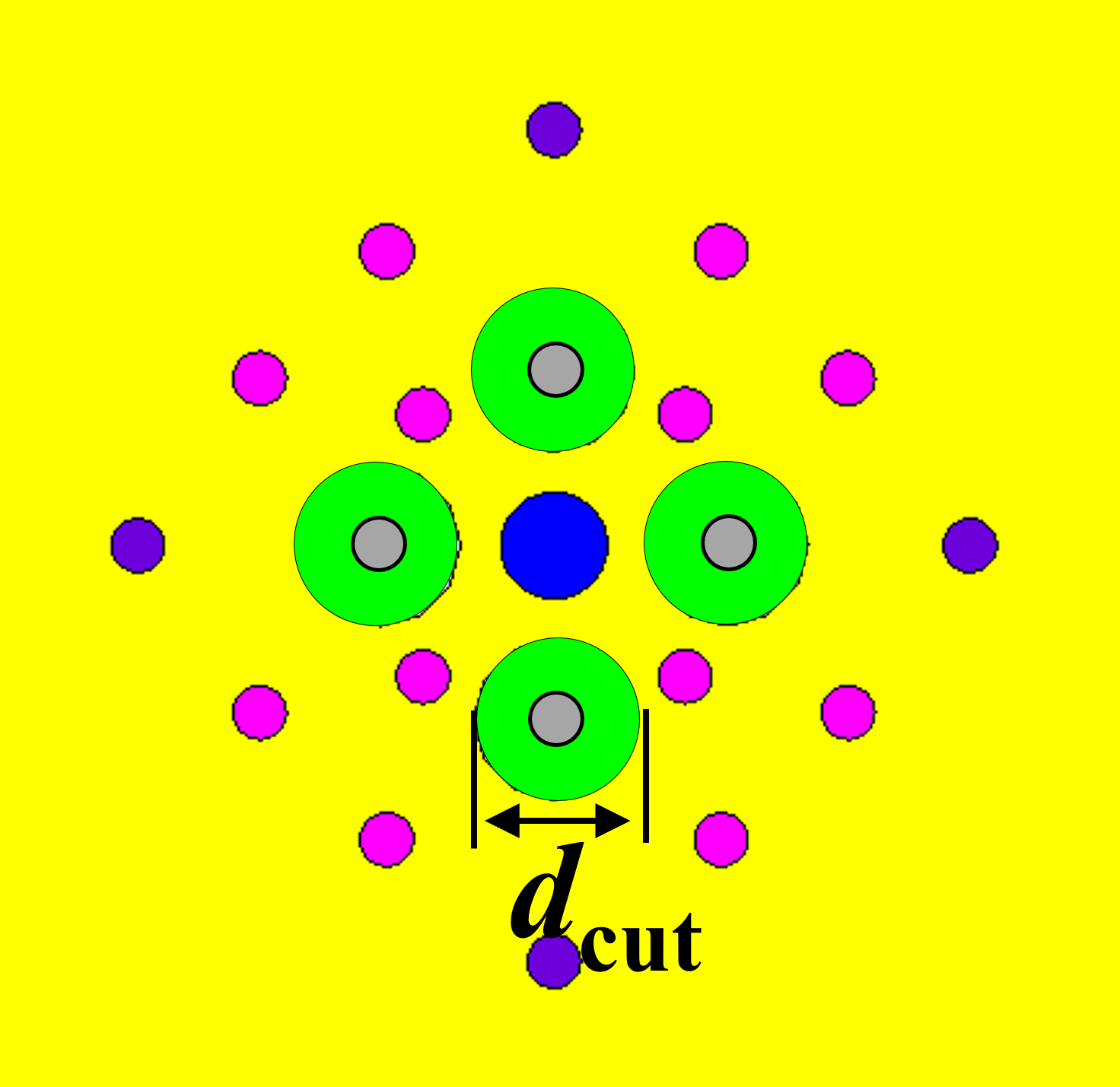}}
\hfill
\subfigure[Layer 7 (L7)]{\includegraphics[width=0.8in]{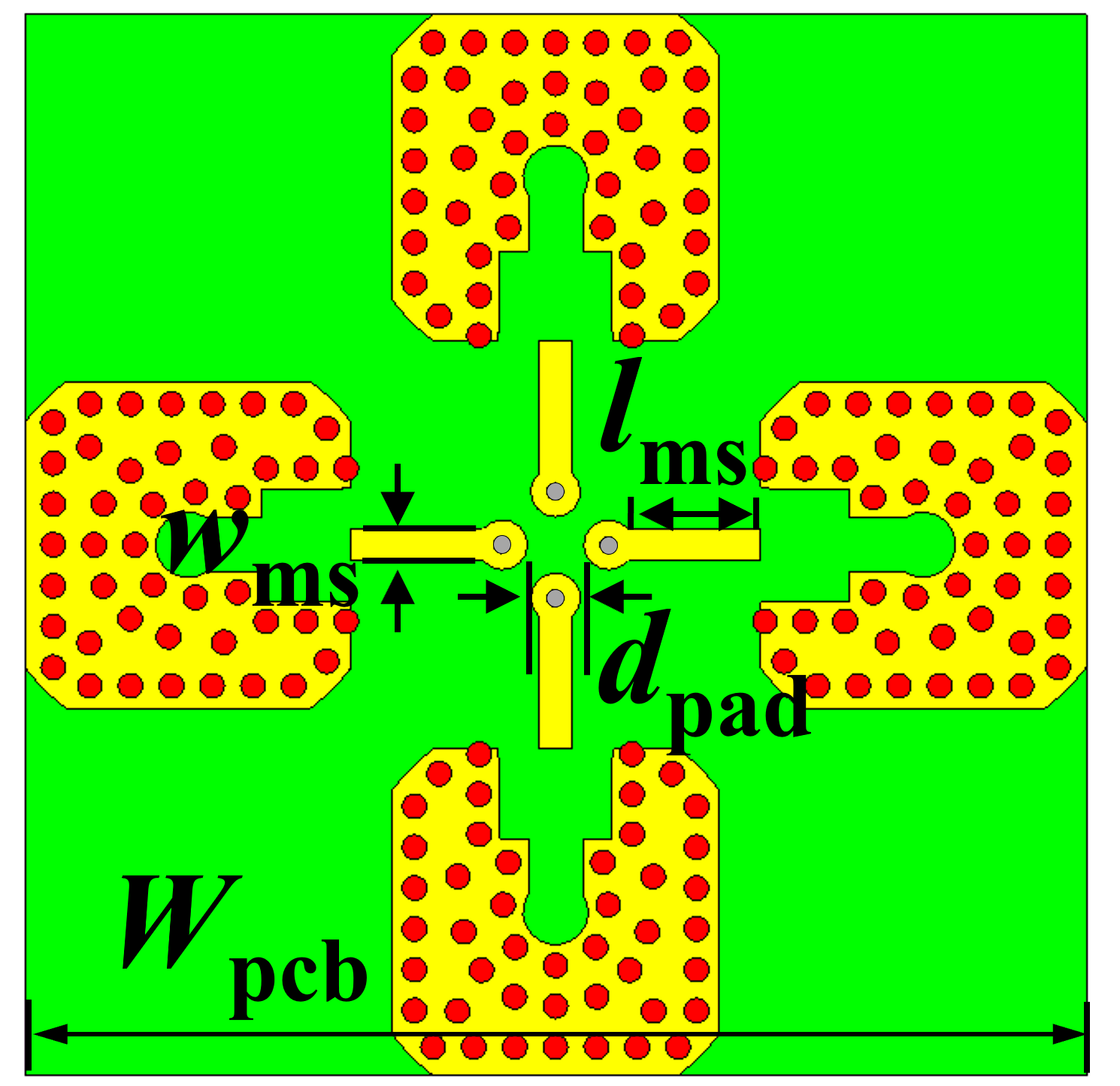}}
\hfill
%\subfigure[]{\includegraphics[width=3.2in]{Plot/exploded1.png}}
\hfill
\caption{Proposed filtering antenna geometry: (a) PCB stack, (b) side view, (c) top view, and (d)-(j) description of metal layers L1-L7.}
\label{configuration}
\end{figure}
\par

\begin{table}[!t]
\caption{Parameters of the Proposed Second-Order Filtering Antenna. Dimensions are in Millimeters}
\begin{center}

\begin{tabular}{m{6.5mm} m{6.5mm} m{6.5mm} m{6.5mm} m{6.5mm} m{6.5mm} m{6.5mm} m{6.5mm}}
\toprule

  \scriptsize  Param.&\scriptsize $W_\mathrm{pat}$ & \scriptsize $W_\mathrm{gnd}$ & \scriptsize $W_\mathrm{cav1}$ &\scriptsize $W_\mathrm{cav2}$ & \scriptsize $W_\mathrm{pcb}$ &\scriptsize $W_\mathrm{array}$  &\scriptsize $w_\mathrm{cor}$ \\
  \hline
  
    \scriptsize  Value &  \scriptsize 3.15  & \scriptsize 2.08 & \scriptsize 10 & \scriptsize 5  & \scriptsize 13  & \scriptsize 35 & \scriptsize 0.4\\

  %  \scriptsize (mm)   & &  &  &  &  &  & \\
 
 %    \midrule
     \hline
     %\midrule
    \scriptsize  Param.& \scriptsize $w_\mathrm{1}$ & \scriptsize $w_\mathrm{2}$ & \scriptsize $w_\mathrm{ms}$ & \scriptsize $w_\mathrm{hp}$ & \scriptsize $w_\mathrm{os}$ & \scriptsize $w_\mathrm{ss}$ & \scriptsize $d_\mathrm{pad}$   \\
  \hline
  
    \scriptsize  Value &  \multirow{1}{*}{\scriptsize 0.37}  & \multirow{1}{*}{\scriptsize 0.55} & \multirow{1}{*}{\scriptsize 0.39} & \multirow{1}{*}{\scriptsize 0.1} & \multirow{1}{*}{\scriptsize 0.1} &  \multirow{1}{*}{\scriptsize 0.1} &  \multirow{1}{*}{\scriptsize 0.61}\\

    %\scriptsize (mm)   & &  &  &  &  &  & \\
 
  %   \midrule
     %\midrule
     \hline
 \scriptsize  Param.& \scriptsize $d_\mathrm{g1}$ & \scriptsize $d_\mathrm{g2}$ & \scriptsize $d_\mathrm{g3}$ & \scriptsize $d_\mathrm{g4}$ & \scriptsize $d_\mathrm{g5}$ & \scriptsize $d_\mathrm{g6}$ &  \scriptsize $d_\mathrm{cut}$ \\
  \hline
  
    \scriptsize  Value &  \multirow{1}{*}{\scriptsize 0.2}  & \multirow{1}{*}{\scriptsize 0.6} & \multirow{1}{*}{\scriptsize 0.4} & \multirow{1}{*}{\scriptsize 0.39} & \multirow{1}{*}{\scriptsize 1.1} &  \multirow{1}{*}{\scriptsize 0.49} &  \multirow{1}{*}{\scriptsize 0.6}\\

    %\scriptsize (mm)   & &  &  &  &  &  & \\
 
%     \midrule
     %\midrule
    \hline
    
 \scriptsize  Param. & \scriptsize $d_\mathrm{hp}$ & \scriptsize $d_\mathrm{p}$ & \scriptsize $d_\mathrm{fp}$ & \scriptsize $d_\mathrm{sp}$ & \scriptsize $l_\mathrm{os}$ & \scriptsize $l_\mathrm{ss}$ & \scriptsize $l_\mathrm{ms}$ \\
  \hline
  
    \scriptsize  Value &  \multirow{1}{*}{\scriptsize 2.11}&  \multirow{1}{*}{\scriptsize 1.3}  & \multirow{1}{*}{\scriptsize 0.4}  & \multirow{1}{*}{\scriptsize 0.4} & \multirow{1}{*}{\scriptsize 1.185}  & \multirow{1}{*}{\scriptsize 0.905} & \multirow{1}{*}{\scriptsize 1.61}  \\

 %   \scriptsize (mm)   & &  &  &  &  &  & \\
 
%     \midrule
     %\midrule
     \hline
 \scriptsize  Param.   &\scriptsize $l_\mathrm{1}$ & \scriptsize $l_\mathrm{2}$ & \scriptsize $l_\mathrm{hp1}$ & \scriptsize $l_\mathrm{hp2}$ & \scriptsize $l_\mathrm{hp3}$ & \scriptsize $l_\mathrm{hp4}$ & \scriptsize $l_\mathrm{hp5}$  \\
  \hline
  
    \scriptsize  Value &  \multirow{1}{*}{\scriptsize 3.69}  & \multirow{1}{*}{\scriptsize 1.3}  & \multirow{1}{*}{\scriptsize 0.35} & \multirow{1}{*}{\scriptsize 0.245}  & \multirow{1}{*}{\scriptsize 0.245} & \multirow{1}{*}{\scriptsize 0.345} &  \multirow{1}{*}{\scriptsize 0.28}   \\

 %   \scriptsize (mm)   & &  &  &  &  &  & \\
 
  %   \midrule
  %    \midrule

  %  \scriptsize  Param. & \scriptsize $l_\mathrm{hp5}$ & \scriptsize $l_\mathrm{os}$ & \scriptsize $l_\mathrm{ss}$ & \scriptsize $l_\mathrm{ms}$ & & &  \\
  % \hline
  
  %   \scriptsize  Value &\multirow{1}{*}{\scriptsize 0.28} &  \multirow{1}{*}{\scriptsize 1.185}  & \multirow{1}{*}{\scriptsize 0.905} & \multirow{1}{*}{\scriptsize 1.61} &  &   & \\

    %\scriptsize (mm)   & &   &  &  &  &  & \\
 
\bottomrule

\end{tabular}
\label{tab1}
\end{center}
\end{table}

\section{Antenna Design}

\subsection{Unit Cell}

Fig.~\ref{configuration} shows the stack and layered view of the proposed differentially driven dual-wideband dual-polarized ME dipole antenna, which consists of a cross-slotted octagonal radiation patch (L1), four squeezed hairpin resonators (L2), an additional lifted ground (L3), four coupled ${\lambda}$/4 open-/short-circuited stub resonators (L4 and L5), a main ground (L6), and microstrip feedlines (L7). Panasonic Megtron 7 laminate and prepreg (bonding material) with $\varepsilon_\mathrm{r}\,=\,3.34$ and $\tan\delta\,=\,0.003$ are used for the stack. Layers L1, L3, and L6--L7 are the same as in the reference wideband ME dipole antenna of \cite{Chen2022}, from which additional information on these layers can be found. 
%radiation or radiating. 
% The electric current distributions at the patch edges contribute to two electric dipole modes while the magnetic current distributions over two orthogonal primary slots generate the magnetic dipole mode. Two secondary slots at the short-circuited ends provide further impedance matching and resonance fine-tuning.
% %It is worth mentioning that the designed ME dipole can be regarded as non-uniform-width polygonal ring patch antenna \cite{bhar01} in which the inner edges of the ring form the aperture serving as a slot antenna. 
% The cross-like lifted ground (LGND) in L3 excites the higher-order mode, tunes the impedance matching and improves the UB gain. The main ground plane of L6 isolates the antenna and feedline substrates and behaves as the reflector for the radiating patch. 

Four plated through-vias (Via 1) act as the feed probes (connecting the slotted patch on L1 to the microstrip feedlines on L7), and they are grouped in two differential pairs at opposite sides. Each differential pair can excite three desired complementary antenna modes for wideband operation.
%An SMPM connector connects the other end of the feedline.
Quasi-coaxial shorting vias (Vias 2--3) encircle the feeding pins to eliminate the parallel-plate mode and to implement the vertical transition. 
Surface wave modes are suppressed with a 1-mm deep air cavity \cite{Li2016ME} around the antenna unit cell (UC).
%(dashed yellow box in Figs.~\ref{configuration}(a) and \ref{Array_layout}(a)) \cite{Li2016ME} around the
%5\(\times\)5-mm$^{2}$ 
%antenna unit cell (UC).
%in the large PCB for the single element prototype. % The air cavity is indicated by the dashed yellow box in Figs.~\ref{configuration}(a) and \ref{Array_layout}(a). 
%Via 4 is a blind laser via while Vias 2--3 are buried vias. 

To create transmission zeros between LB and UB, two resonator types are added to the reference design. Four hairpin resonators symmetrically below the slot along E- and H-planes couple magnetically to the slot ends. With four $\lambda/4$ coupled open-/short-circuited stub resonators, one end of both stubs connects directly to the feed probes (within the quasi-coaxial line), and a shorting via (Via 5) terminates the other end of the short-circuited stub to the main ground (L6).
%Details of the layers and dimensions are given in Fig.~\ref{configuration}(b)-(h). 
%The antenna is symmetric in $xz$ and $yz$ planes. 
%The radiating patch with a step-impedance cross slot shown in Fig.~\ref{configuration}(b) is printed on the top layer and fed by the differential pairs (Ports 1--2) along the symmetry planes ($xz$ and $yz$ planes).

Fig.~\ref{configuration}(e) shows that the hairpin resonators are squeezed to miniaturize the $\lambda/2$ resonator and reduce its footprint. 
%As shown in Fig.~\ref{configuration}(d), the vertical transition incorporates shorting Vias 2--3 encompassing the feeding pins to create a quasi-coaxial line. 
%Note that the feeding probes serve as part of the bandstop filter and act as non-redundant connecting lines (unit cells in filter) which will be further The lifted ground plane is in favor of the reduction and cancellation of the large capacitive reactance for improving the impedance matching. 
The open-circuited stubs of L4 and short-circuited stubs of L5 are broadside-coupled for high $Q$, using an embedded microstrip structure as shown in Figs.~\ref{configuration}(g)--(h). 
%Combining the open- and short-circuited stubs creates the desired ${\lambda}$/4 coupled stub resonator that uses an embedded microstrip structure. 
The footprints of both resonator types overlap under the patch for almost the same element size as without the resonators to miniaturize the design. 
%In Fig.~\ref{configuration}(g), four concentric circular cutoffs with radius of 0.3\,mm are made in the main ground plane for the feeding probes. Both the SMPM connector and microstrip feedlines shown in Fig.~\ref{configuration}(h) have $Z_0=$50\,$\Omega$. 
Table~\ref{tab1} lists the optimized design parameters.

\begin{figure}[!t]
\hfill
\subfigure[Top view]{\includegraphics[width=1.59in]{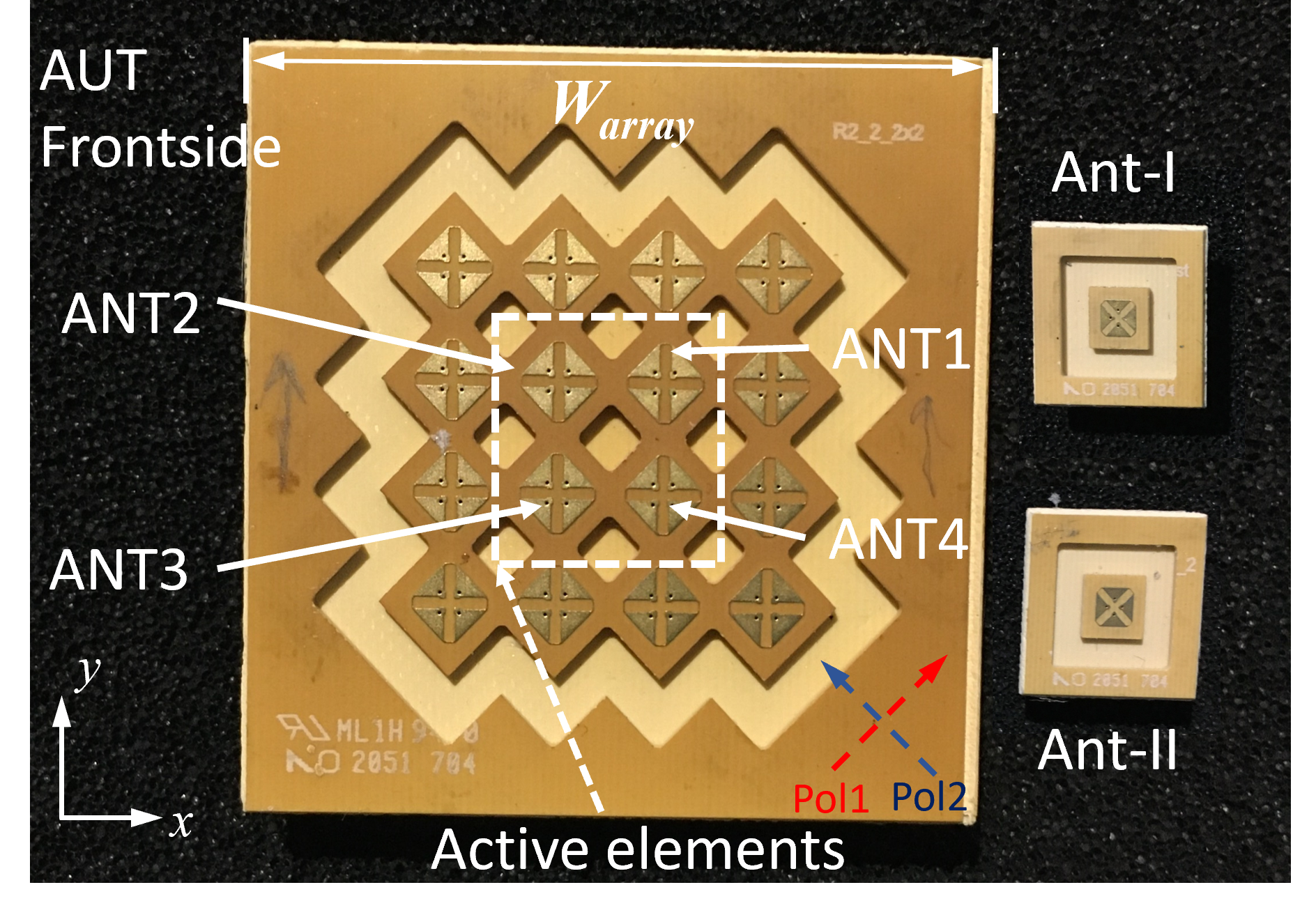}}
\hfill
\subfigure[Bottom view]{\includegraphics[width=1.54in]{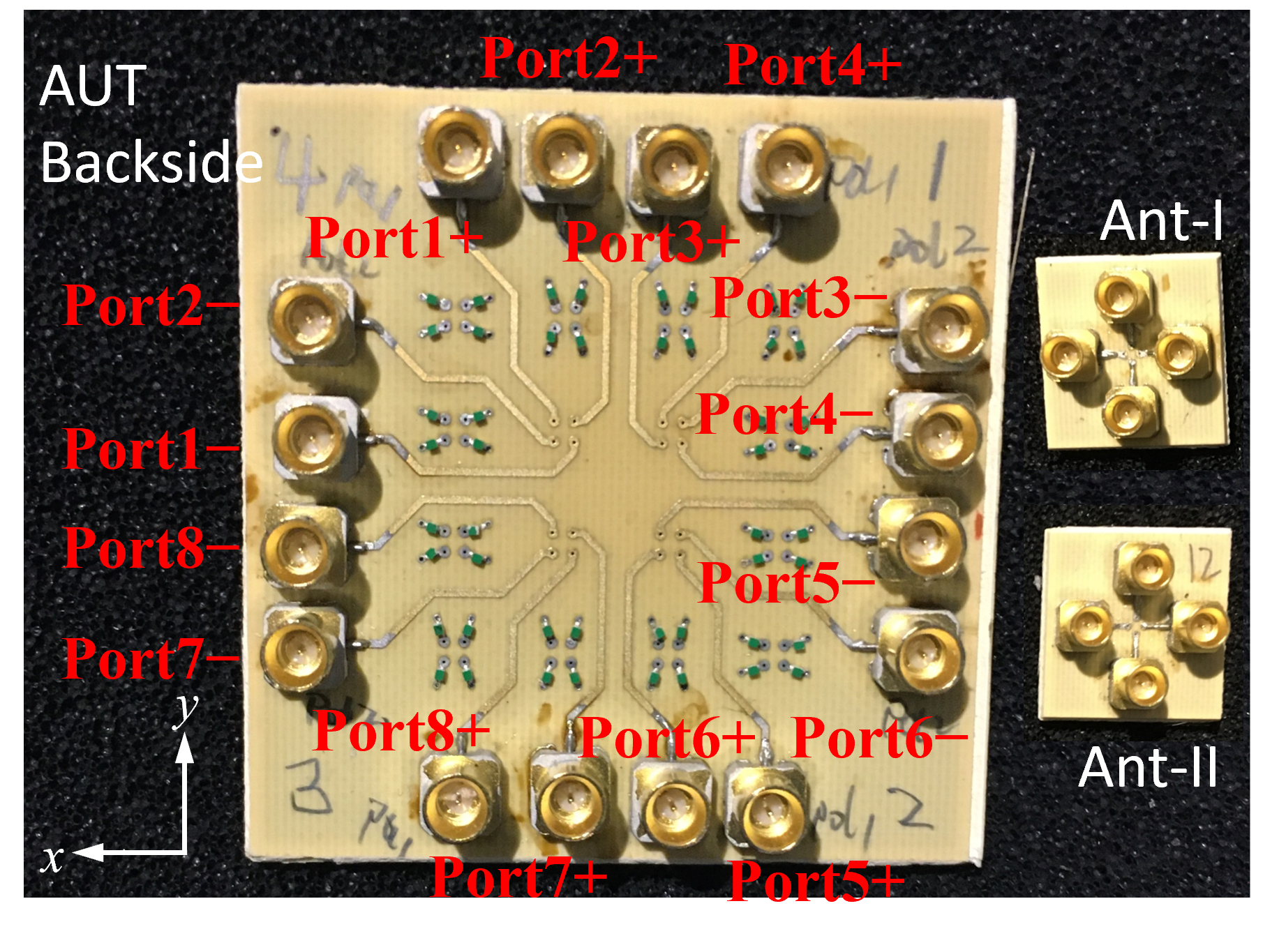}}
% \centering
% \subfigure[Unit cell (right) and 2$\times$2 array (left)]{\includegraphics[scale=0.27]{Plot/prototype2.png}}
\caption{Prototype of the fabricated 2$\times$2 array and unit cells.}
  \label{Array_layout}
\end{figure}

\begin{figure}[!t]
\centering
\includegraphics[scale=0.251]{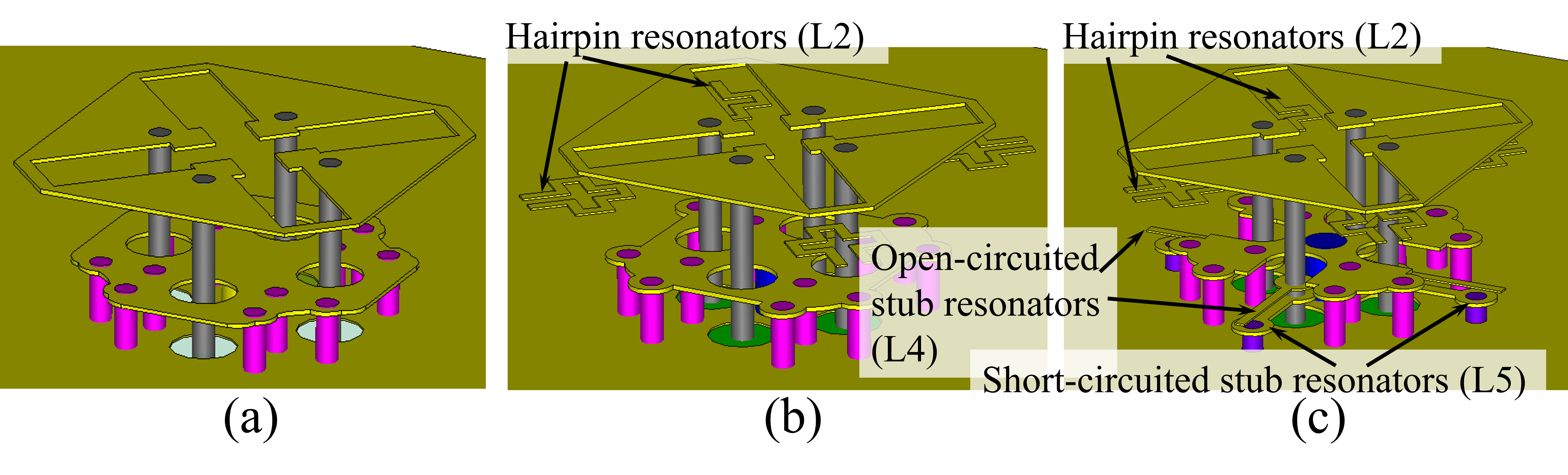}
\caption{
Filtering antenna evolution: (a) reference wideband antenna \cite{Chen2022}; (b) first-order filtering (Ant-I); (c) second-order filtering (Ant-II, proposed).
%Evolution of the proposed filtering antenna: (a) reference wideband antenna \cite{Chen2022}; (b) Ant-I with first-order bandstop filtering; (c) Ant-II (the proposed) with second-order bandstop filtering.%\red{figure (b) and (c) seem have worse resolution than (a) when I zoom in}. \blue{This might be the case; the result could be improved by scaling the originals without the text/arrows.}
}
\label{evolution}
\end{figure}

\begin{figure}[!t]
\hfill
\subfigure[]{\includegraphics[scale=0.27]{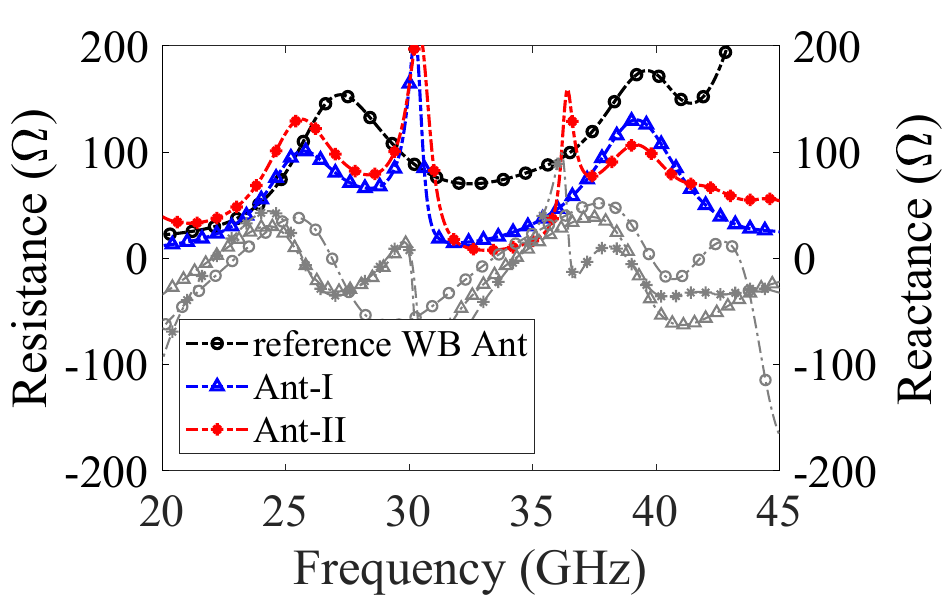}}
\hfill
\subfigure[]{\includegraphics[scale=0.27]{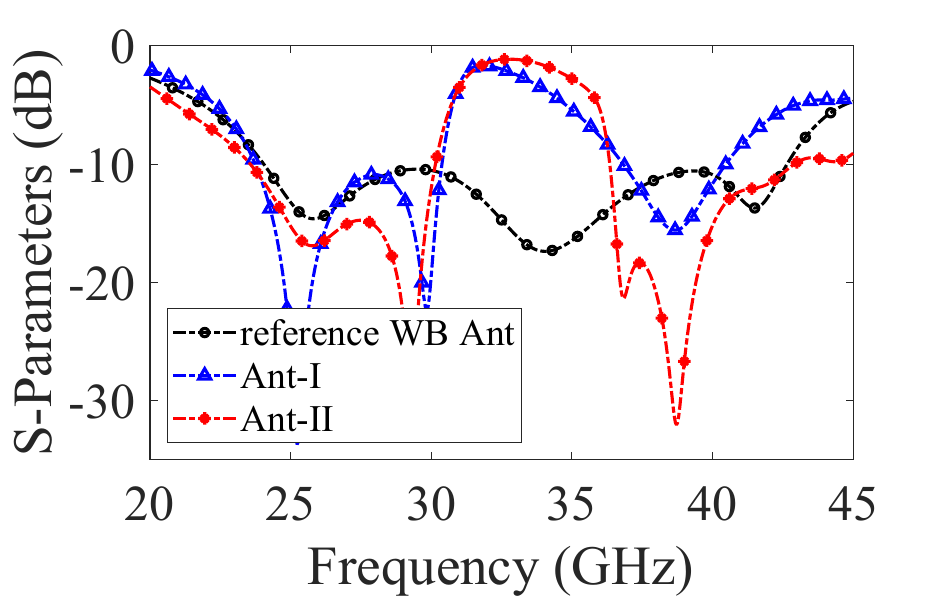}}
\hfill
\caption{(a) Input impedance and (b) reflection coefficients of the proposed second-order band-notched antenna (Ant-II) with comparison to reference wideband antenna and the first-order band-notched antenna (Ant-I).% \red{Add the dash to legend (Ant-I and Ant-II)}
}
  \label{evolution_S_para}
\end{figure}

\subsection{2$\times$2 Array}
%\IEEEpubidadjcol

To verify the array performance of the proposed unit cell, a planar 2\(\times\)2 array with \(\pm\)45\(\degree\) slanted polarization is built (see Fig. \ref{Array_layout}). Four active elements in the middle of array are fed with individual differential pairs instead of a corporate feed network to see the isolation between adjacent and diagonal elements. The unit cell spacing is 0.57${\lambda_{0}}$ at 32\,GHz (trade-off between LB and UB operation). 
% The fabricated prototype of array and single elements are shown in Fig.2 
%Also, the spacing compromizes between grating lobe performance and the room needed to accommodate 16 SMPM connectors and feed lines. The unit cells in the 2$\times$2 array follow the single-element design with some further parameter optimization for matching purposes.

The array has 12 dummy elements terminated to 50\,$\Omega$ around the inner four active elements to overcome the edge effects \cite{Gu2019} of the 2$\times$2 array on a large ground plane. The air cavity around the dummy elements and between active elements reduces the substrate mode effect.  Fig.~\ref{Array_layout}(b) shows the rotationally symmetric feed network routing for four differential-fed dual-polarized active elements.%; all 16 feed lines are rotationally symmetric. 

%In contrast with the single ended antenna, amplitude and phase imbalance may occur due to the in each differential pair in case of a differentially-fed antenna.

% \begin{figure}[!t]
% \centering
% \includegraphics[scale=1]{IEEEtran/Plot/antenna_evolution_v3.pdf}
% \caption{Evolution of the proposed filtering antenna: (a) reference wideband antenna \cite{Chen2022}; (b) Ant-I with first-order bandstop filtering; (c) Ant-II (the proposed) with second-order bandstop filtering.%\red{figure (b) and (c) seem have worse resolution than (a) when I zoom in}. \blue{This might be the case; the result could be improved by scaling the originals without the text/arrows.}
% }
% \label{evolution}
% \end{figure}

% \begin{figure}[!t]
% \hfill
% \subfigure[]{\includegraphics[scale=0.27]{Plot/evolution_Zin3.eps}}
% \hfill
% \subfigure[]{\includegraphics[scale=0.27]{Plot/evolution_S_para3.eps}}
% \hfill
% \caption{(a) Input impedance and (b) reflection coefficients of the proposed second-order band-notched antenna (Ant-II) with comparison to reference wideband antenna and the first-order band-notched antenna (Ant-I).}
%   \label{evolution_S_para}
% \end{figure}

\section{Operation Principle}\label{principle}

\begin{figure*}[!t]
\hfil
\subfigure[]{\includegraphics[scale=0.57]{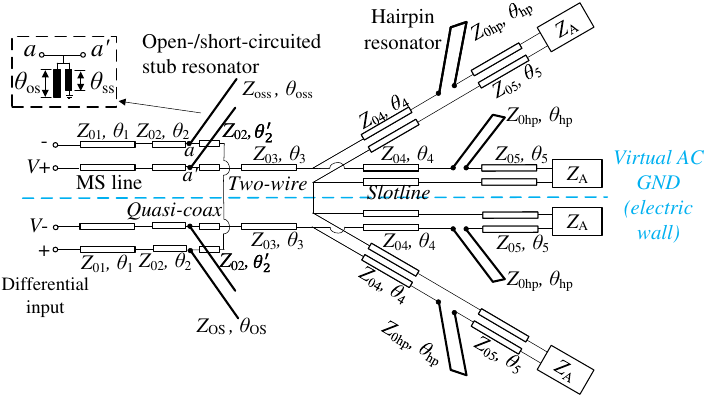}
}
\hfil
\subfigure[]{\includegraphics[scale=0.57]{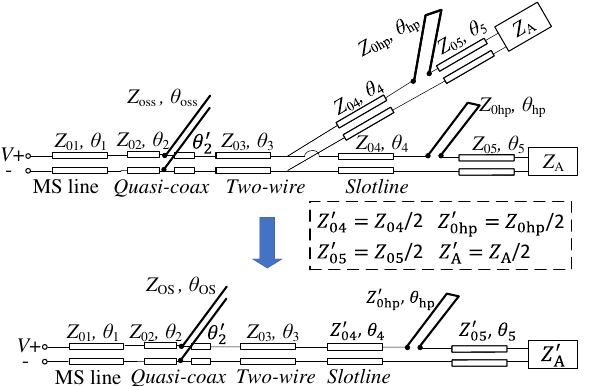}
}
\hfil
\subfigure[]{\includegraphics[scale=0.57]{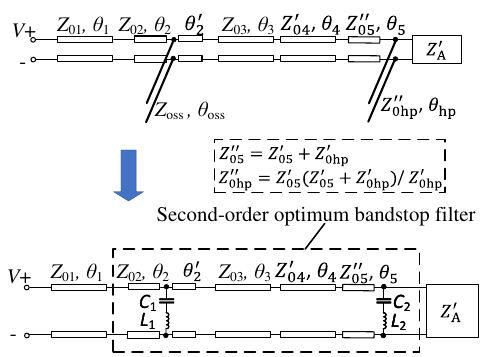}
}
\hfil
\caption{
Equivalent circuit model for the proposed filtering antenna: (a) original circuit, (b) simplified circuit (bisection), and (c) the resultant equivalent circuit for the second-order optimum bandstop filter after applying Kuroda's identities. The circuit model is used to qualitatively explain the antenna operation.
%Z0hp is dependent on the impedance inverter
%(a) Side view and (b) top view of the equivalent circuit components in the proposed antenna; (c) original equivalent circuit model, (d) simplified equivalent circuit model (bisection), and (e) the resultant equivalent circuit model for the second-order optimum bandstop filter after applying Kuroda's identities. %\red{I modified different versions and font size in the present circuit model seems the largest possible one I can make otherwise it will block the line model. Still there are some spacing due to (a) and (b), do you think we can make (a) and (b) single-column version and (c)-(e) double-column one. Then we can fully use the space. That means (a) and (b) will be a separate figure e.g. Fig. 5 and (c)-(e) will become Fig. 6. Or we can try to modify (a) and (b)}
}
\label{ECM}
\end{figure*}
\par
\subsection{Evolution of the Second-Order Bandstop Antenna}\label{evo}

To understand the operating principle of the proposed second-order bandstop antenna from the reference wideband ME dipole \cite{Chen2022}, Fig.~\ref{evolution} presents the antenna evolution step by step. %he process of the antenna developmentfrom fundamental wideband antenna to the second-order bandstop antenna is given in Fig. \ref{evolution}, 
Fig.~\ref{evolution_S_para} shows the corresponding input impedance and reflection coefficient.
The first modification (Ant-I) adds a set of hairpin resonators to create the first transmission zero. 
%These resonators are below the antenna patch and couple magnetically to the slot ends. 
%(open end facing away from the center of the crossed slot).
A second modification (Ant-II) adds coupled stub resonators to create a second transmission zero, to improve stopband rejection and band-edge selectivity, and to control the stopband BW. The coupled stubs also create passband transmission poles whose locations can be controlled to enhance UB matching.
%In L3, the lifted ground of Ant-II is slightly modified for improved matching when adding the stub resonators.

As seen in Fig.~\ref{evolution_S_para}(a), the differential input resistance of the reference wideband antenna fluctuates around 100\,$\Omega$. The --10-dB bandwidth is 24.0--42.6\,GHz %(shown in Fig.~\ref{evolution_S_para}(a)), 
which does not cover the desired n259 band (up to 43.5\,GHz). After adding the hairpin resonators, the resistance is close to zero around 32\,GHz, creating the first transmission zero (radiation null). Meanwhile, the UB matching deteriorates, reaching only up to 40.5\,GHz
%at the upper bound of UB 
despite good LB matching. Also, the upper stopband-edge selectivity is low.

Coupled ${\lambda}$/4 open-/short-circuited stub resonators are used as the second resonator type in Ant-II. They have high $Q$, compact size, and independent control of the transmission zeros and poles \cite{Chuang2012}. Traditional ${\lambda}$/4 open-/short-circuited stub resonators of \cite{Yim2005} have low inductance and low $Q$, which is not suitable for narrow-band bandstop filters. In \cite{Chuang2012}, the open- and short-circuited stubs are folded to couple them to each other. This increases inductance, reduces the form factor, and increases $Q$. Transmission zeros and poles depend on the electrical length of open- and short-circuited stubs, respectively.
%Equal stub length gives two symmetrical transmission poles at either side of the transmission zero.
Varying the length of short-circuited stub with fixed length of open-circuited stub causes different frequency shift at the upper and lower transmission poles for an extra degree of freedom in passband matching.
However, the co-planar layout of the coupled stub resonators is asymmetric over the patch symmetry plane due to edge-coupling, which degrades port-to-port isolation and reduces XPD. PCB fabrication rules limit the possible line spacing and width, which can complicate designing narrow stubs and narrow-band filters. 

For structural symmetry,
we propose a symmetric broadside-coupled stub structure with the open- and short-circuited stubs at two successive layers (L4 and L5), as shown in Fig.~\ref{configuration}(e)--(f). The coupling between two stub types can be altered by changing the laminate thickness between the layers. %, which is easier to implement. 
After adding the stub
%${\lambda}$/4 broadside-coupled open-/short-circuited stub 
resonators, the resistance of Ant-II has a second transmission zero (radiation null) near 35\,GHz. The UB bandwidth improves to cover the desired bands, while both the LB matching and the first transmission zero remain stable (see Fig. \ref{evolution_S_para}). This improves the upper band-edge selectivity at stopband. Also, the location of the second resonator below the lifted ground has low coupling to the hairpin resonators due to the increased physical separation and different ground references to keep them working independently.

\subsection{Filter Synthesis and Equivalent Circuit Model}

Figs.~\ref{configuration}(b)--(c) show that each part of the feeding structure can be modeled as transmission line sections. From the differential port to the patch, the input signal sees a section of microstrip, quasi-coaxial, two-wire and slotline transmission lines. Their characteristic impedances and electrical lengths are given in the original equivalent circuit model of Fig.~\ref{ECM}(a).
%($Z_\mathrm{01}$, $\theta_\mathrm{1}$), ($Z_\mathrm{02}$, $\theta_\mathrm{2}$+$\theta_\mathrm{2}^{\prime}$), ($Z_\mathrm{03}$, $\theta_\mathrm{3}$), ($Z_\mathrm{04}$, $\theta_\mathrm{4}$) and ($Z_\mathrm{05}$, $\theta_\mathrm{5}$).
%\red{(can we just say their characteristic impedance and electrical length are marked in Fig.~\ref{ECM_ant}? And we can describe the equations in the figure instead of in the paragraphs, then we can save some lines in this subsections.)}
%the hairpin resonators use the slotted patch as reference plane while the stub resonators use the main GND plane as reference plane.
%Fig.~\ref{ECM} depicts the original equivalent circuit model. 
The inset shows the schematic of the coupled open-/short-circuited stub.
%, where $\theta_\mathrm{os}$ and $\theta_\mathrm{ss}$ are the electrical length of the open- and short circuited stubs, respectively. 
Since the used ${\lambda}$/4 coupled open-/short-circuited stub resonator acts as a series $LC$ resonator at the center frequency \cite{Chuang2012}, an open stub is used for convenience in the equivalent circuit model. The ${\lambda}$/2 hairpin resonator is modeled as a short-circuited stub in series as indicated in \cite{hong2004microstrip}. 
%The characteristic impedance and electrical length of the open- and short-circuited stubs are ($Z_\mathrm{oss}$, $\theta_\mathrm{oss}$) and ($Z_\mathrm{0hp}$, $\theta_\mathrm{hp}$), respectively. 
$Z_\mathrm{A}$ is the terminating impedance (the input impedance of the reference ME dipole under symmetry condition). A circuit model for $Z_\mathrm{A}$ with three parallel-coupled serial $RLC$ resonators is shown in \cite{Chen2022}.

Due to odd-mode excitation of a symmetrical network and the fact that a symmetry plane accounts for a short circuit of the bisection (electric wall), the original equivalent circuit is developed into a simpler one in Fig.~\ref{ECM}(b) in accordance with \cite{LiuSIR2016, Chen2022}. Characteristic impedance values are halved as described in Fig.~\ref{ECM}(b).
%so that $Z_\mathrm{04}^{\prime}$=$Z_\mathrm{04}$/2, $Z_\mathrm{05}^{\prime}$=$Z_\mathrm{05}$/2, $Z_\mathrm{0hp}^{\prime}$=$Z_\mathrm{0hp}$/2, and $Z_\mathrm{A}^{\prime}$=$Z_\mathrm{A}$/2. 
Applying Kuroda’s identities gives the resultant equivalent circuit of Fig.~\ref{ECM}(c),
%where $Z_\mathrm{05}^{\prime\prime}$ = $Z_\mathrm{05}^{\prime}$+$Z_\mathrm{0hp}^{\prime}$ and $Z_\mathrm{0hp}^{\prime\prime}$ = $Z_\mathrm{05}^{\prime}$($Z_\mathrm{05}^{\prime}$+$Z_\mathrm{0hp}^{\prime}$)/$Z_\mathrm{0hp}^{\prime}$ \cite{hong2004microstrip}. 
corresponding to a typical second-order (two-pole) bandstop filter. The shunt $\lambda/4$ open-circuited stubs connect to unit elements whose length are $\lambda/4$ at the stopband center frequency in a traditional bandstop filter with open-circuited stubs. Hence, the connecting lines ($\theta_\mathrm{2}^{\prime}$, $\theta_\mathrm{3}$ and $\theta_\mathrm{4}$) between two stubs can be modeled as filter unit cells. 

For better band-edge attenuation, the unit cell is made non-redundant to effectively make it an open-circuited stub \cite{hong2004microstrip}, differing from the redundant one in a traditional bandstop filter. This makes the feeding probes in the proposed antenna part of the filter element. In narrow-band filters, the stubs in a redundant design can become very narrow and hard to implement. Fabrication limits are more relaxed for a non-redundant design. Fewer stubs can produce a steeper stopband edge in a non-redundant unit cell than in a redundant one. A non-redundant design can be achieved by adjusting design parameters (e.g., terminating impedance, electrical length and line impedance) of both resonators and connecting lines shown in Fig.~\ref{ECM}(c).%, and the effect of key design parameters will be discussed in Section~\ref{array}.

\section{Simulation and Measurement Results}
%To validate the simulations, the 
The single element and 2\(\times\)2 array of Fig.~\ref{Array_layout} are fabricated. $S$-parameters of the antenna ports are measured using a Keysight N5247B PNA-X network analyzer. %, with each port connected to an SMPM connector on the antenna PCB through a coaxial cable. A thru–reflect–line (TRL) calibration kit\cite{Engen1979TRL} is used since the connection between SMPM connectors and coaxial cable is not a standard interface for the PNA. The 
Radiation patterns of single and array elements are measured using the approach of \cite{Chen2022}.% (not shown here for brevity).

% With the TRL calibration technique, the reference plane of the port is moved closer to the antenna element, and the effects of the connector and feed line are deembedded. Time-domain gating was employed to the measured $S$-parameters as the long connection path and discontinuity from antenna port to PNA port via connectors, adapters and cables causes mismatch and a ringing effect \cite{Yin2019Epatch}. Moreover, the differential mode $S$-parameters are calculated based on the technique described in \cite{Sun2021}.

%\red{the measured frequency is up to 40GHz due to the limitation of facility.}

%The measurement setup for the radiation patterns is the same as that used in \cite{Chen2022}, which is not shown here for brevity. \red{The radiation pattern of the single-element prototype and unit cell in the array were measured directly with a Marki BAL0067 mm-Wave balun. For the array, the pattern of the array prototype was achieved by sequentially exciting each unit cell fed by coaxial cables with the remaining unit cells terminated to 50\,$\Omega$. Thereafter, the four individually measured patterns were computationally combined to obtain the array pattern at boresight direction with equal phase and amplitude weighting in MATLAB as implemented in \cite{Nuutti2021, Sun2021Dualband}.}

\subsection{Single Element}

% \begin{figure}[!t]
% \centering
% \includegraphics[scale=0.83]{Plot/chen9.eps}
% \caption{Simulated and measured $S$-parameters for the single element.}
% \label{Fig. 9.}
% \end{figure}
% \par

\begin{figure}[!t]
\hfill
\subfigure[First-order bandstop filtering.]{\includegraphics[scale=0.258]{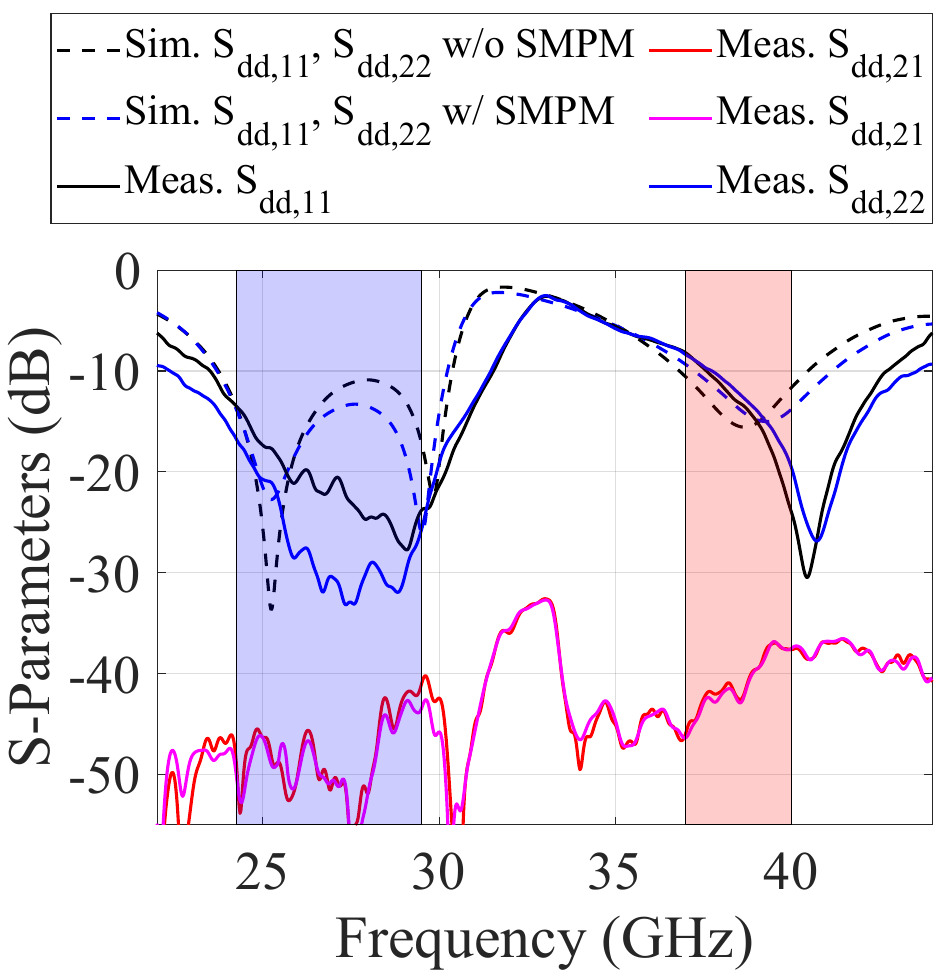}}
\hfill
\subfigure[Second-order bandstop filtering.]{\includegraphics[scale=0.258]{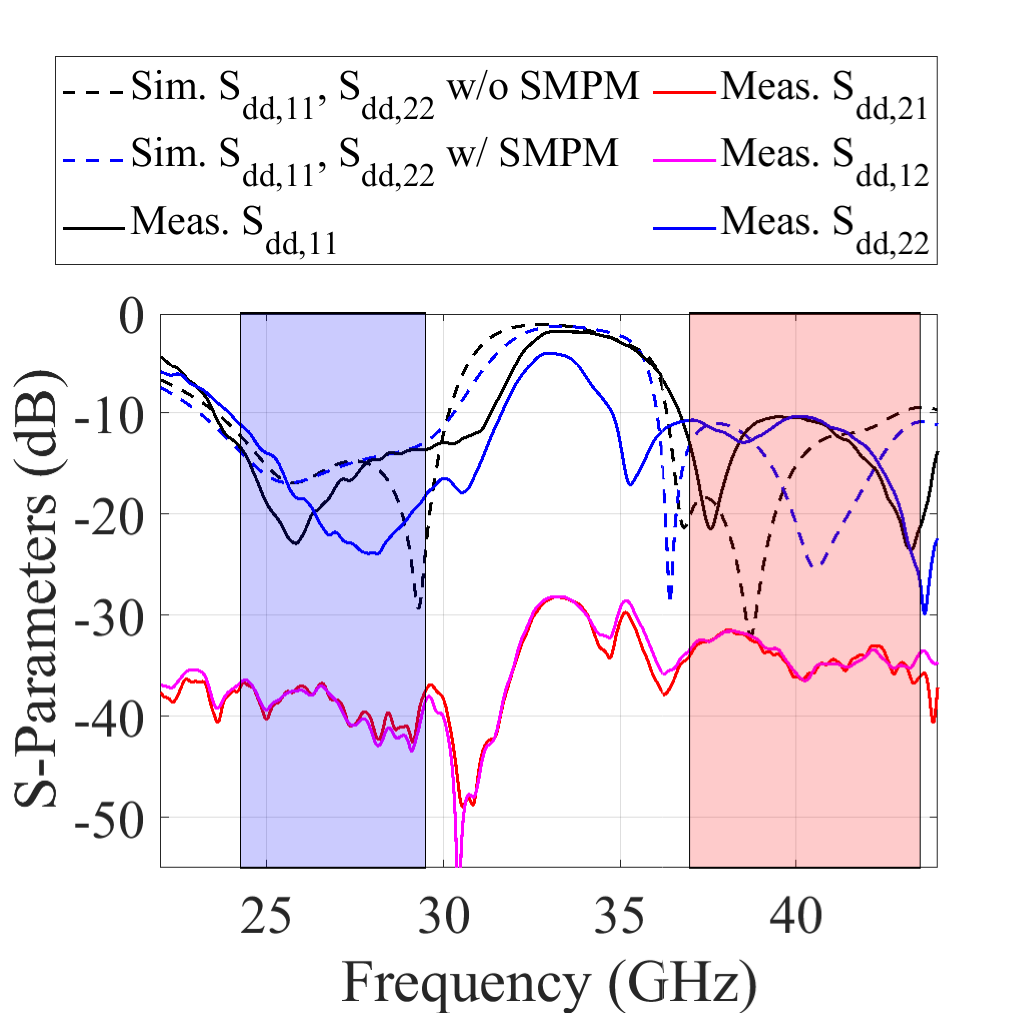}}
\hfill
\caption{$S$-parameters (sim. and meas.) of a single-element filtering antenna. %\red{There is some extra white space around the figures (compare the level of the upper edges of Figs. 6 and 8).}\blue{I use Inkscape to cut the margin of the figure from Matlab}
} %\red{any need to increase the line width?}. \blue{The lines could be made a bit thicker. Also, if you remove the word 'connector' from the legend (i.e. have w/o SMPM and w/ SMPM), it might be possible to have slightly larger font. To me, the x and y axes look alright.}}
  \label{S_para_UC}
\end{figure}

% \begin{figure}[!t]
%   \centering
%   \begin{tabular}[b]{c}
%     \includegraphics[scale=0.15]{Plot/S_para_1st1.eps} \\
%     \footnotesize (a) First-order bandstop filtering.
%     \end{tabular} \qquad
%   \begin{tabular}[b]{c}
%     \includegraphics[scale=0.15]{Plot/S_para_2_2_1.eps} \\
%     \footnotesize (b) Second-order bandstop filtering.
%   \end{tabular}
%   \caption{Effect of filtering order on simulated and measured single-element $S$-parameters. \red{can we try to merge (a) and (b) in one plot or make them in one row?} \blue{One row could work. In that case, I would place the label box on top of the 1x2 sublot array to keep it more legible.}}
%   %\caption{Illustration of (a) simulated impedance matching and (b) measured detector output power. }
%   \label{S_para}
% \end{figure}

% \begin{figure}[!t]
% \centering
% \includegraphics[scale=0.55]{Plot/S_para_1st.eps}
% \caption{Simulated and measured $S$-parameters for the single element with first order filtering.}
% \label{S_para_1st}
% \end{figure}
% \par

% \begin{figure}[!t]
% \centering
% \includegraphics[scale=0.55]{Plot/S_para_2_2.eps}
% \caption{Simulated and measured $S$-parameters for the single element with second order filtering.}
% \label{S_para_2_2}
% \end{figure}
% \par

%  \begin{figure}[!t]
% \centering
% \includegraphics[scale=0.65]{Plot/chen10.eps}
% \caption{Calculated normalized modal weighting coefficients for excitation with the differential feed.}
% \label{Fig. 10.}
% \end{figure}

The simulated and measured $S$-parameters for Ant-I and Ant-II (single element) are presented in Fig.~\ref{S_para_UC}, with a decent agreement between the results. In simulations (Fig.~\ref{S_para_UC}(a)), Ant-I shows full LB coverage for 5G n257, n258 and n261 bands (24.25--29.5\,GHz) while it provides partial UB coverage (n260, 37--40\,GHz only) with a slight UB resonance offset in the measurements. Ant-II obtains a --10-dB matching level across the desired LB and UB range, % depicted by the light blue and red shadowed areas as shown in Fig.~\ref{S_para}(b), 
covering an overlapped bandwidth of 23.9--31.6\,GHz (27.7\(\%\)) at LB and 36.7--45.0\,GHz (>20.3 \(\%\) as frequencies above 45\,GHz were not studied) at UB for two orthogonal polarizations. Comparing the results of Ant-I and Ant-II shows that the second transmission zero enhances band-edge selectivity at the upper bound of the stopband. %Discrepancies between the simulations and measurements are likely caused by fabrication tolerance. %Also, it is noticed that the SMPM connector slightly affects the matching due to soldering inaccuracies and the discontinuity in the inner conductor of the SMPM connector to microstrip transition.

The simulated mutual coupling is not shown as two orthogonal polarizations with symmetric differential feed have ideally infinite isolation. %Even though fabrication tolerances may cause some asymmetry in the prototype, 
The Ant-I/Ant-II achieve better than 40.2\,dB/36.4\,dB and 37.3\,dB/31.6\,dB of isolation at LB and HB, respectively. Additionally, the simulated total efficiencies are above 81.5\(\%\) and 82.8\(\%\) across the frequencies of interest at LB and UB, respectively. 

\begin{figure}[!t]
\centering
\includegraphics[scale=0.43]{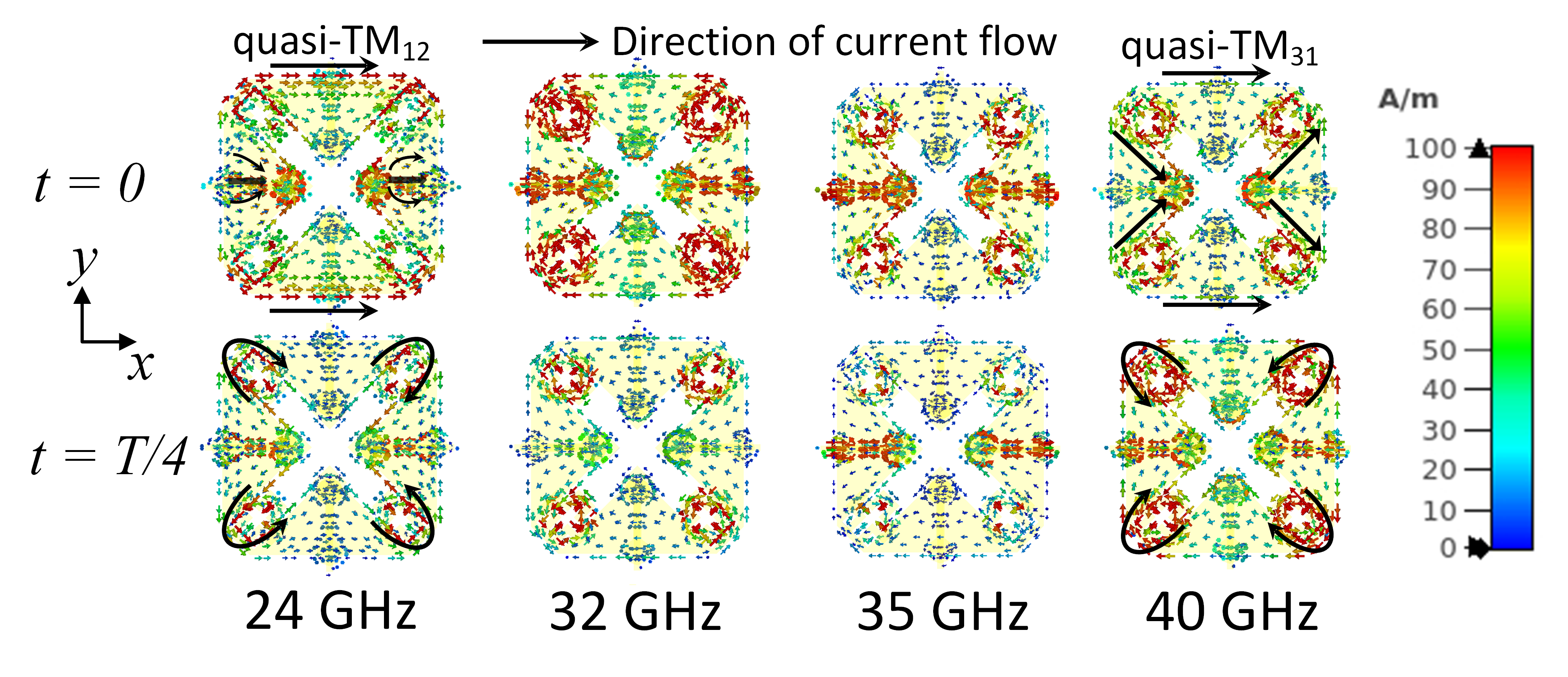}
\caption{Surface current distributions at different time instants. %The thick black arrows indicate the direction of current flow. \red{Try putting a black arrow with the label \enquote{Current flow} between the quasi-TM labels. This could save one row from the label.} %\blue{how about now} \red{Now it's OK. The scale could be made a bit bigger since there is anyway space, but that is a minor thing.}
}
\label{Current}
\end{figure}
\par

Fig.~\ref{Current} shows the current distributions of the proposed band-notched antenna.
%It is clearly indicated that the operation of the antenna has three resonant modes i.e., i.e., quasi-TM$_{12}$, slot (aperture), and quasi-TM$_{31}$ modes at passband.
The quasi-TM$_{12}$ and slot modes are excited at LB passband while the quasi-TM$_{31}$ and slot modes are driven at UB passband. The alternating excitation between electric dipole modes and magnetic dipole (slot) mode with 90\(\degree\) phase difference shows that the operation of reference ME dipole in \cite{Chen2022} is not disturbed by the added resonators. 
%the operation of the band-notched ME dipole, the current distributions of the proposed antenna are shown in Figs.~\ref{Current}. 
% As discussed in \cite{Chen2022}, the reference wideband ME dipole antenna has three resonant modes i.e., quasi-TM$_{12}$, slot (aperture), and quasi-TM$_{31}$ modes. After adding two sets of resonators into the reference antenna, the three modes are kept without disturbance. In other words, the operation of the proposed antenna still remain these three modes with two introduced transmission zeros between the LB and UB to suppress the unwanted radiation. At time instants $t\!=\!0$, Figs.~\ref{Current}(a) and (d) illustrate maximum driven electric current distributions at the top patch in passband with the differential probe feeding, showing the two desired electric dipole modes i.e. quasi-TM$_{12}$ at LB and quasi-TM$_{31}$ modes at UB are excited. At time instants $t\!=\!T/4$, the current distributions are mainly concentrated around the cross-slot ends, indicating that the resonant modes for both LB and UB switch to slot mode (magnetic dipole) which is in favor of the operation of ME dipole antenna since the alternative excitation between electric dipole modes and magnetic dipole mode with 90\(\degree\) phase difference. 
At stopband, the surface currents concentrate on the squeezed hairpin resonators at 32\,GHz and coupled open/short-circuited stub resonators at 35\,GHz. Thus the energy is stored in the resonators instead of radiating, thereby generating two radiation nulls. 

\begin{figure}[!t]
\centering
\includegraphics[scale=0.53]{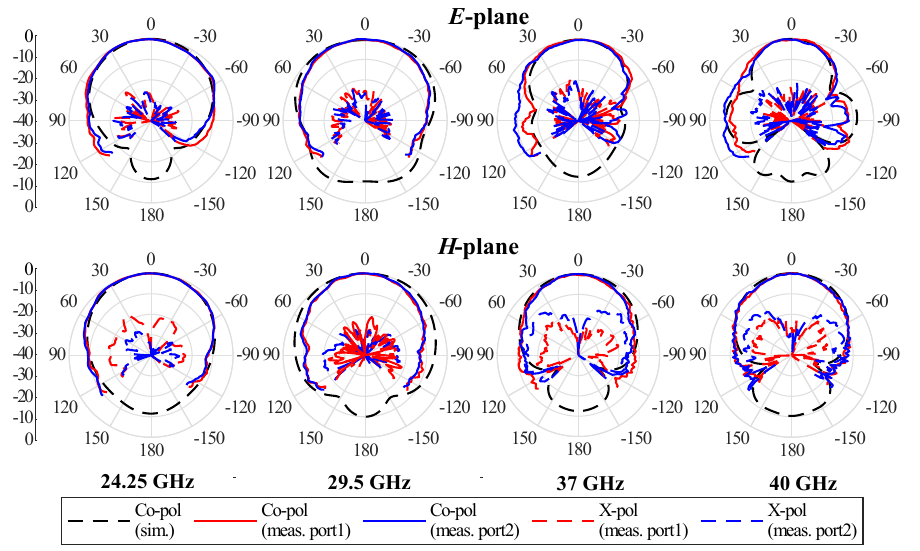}
\caption{Simulated and measured single-element radiation patterns with second-order filtering at 24.25\,GHz, 29.5\,GHz, 37\,GHz and 40\,GHz.}
\label{rad_UC_2_2}
\end{figure}
\par

\begin{figure}[!t]
\subfigure[Simulation.]{\includegraphics[width=1.72in]{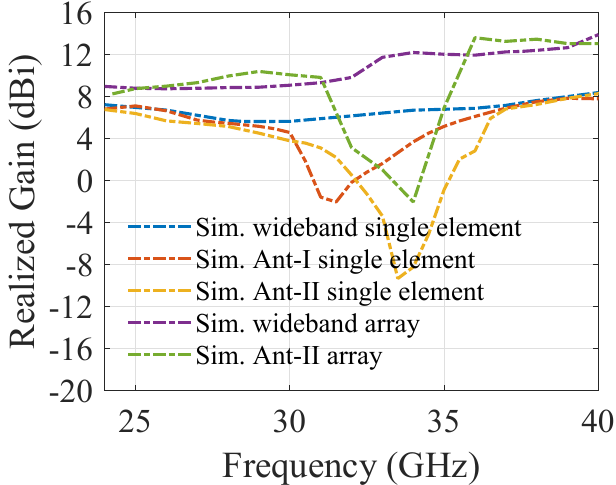}}~
\subfigure[Measurement.]{\includegraphics[width=1.72in]{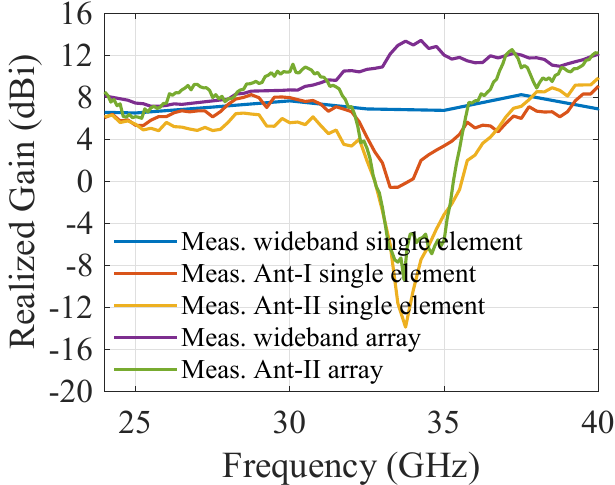}}
\caption{Comparison of (a) simulated and (b) measured realized gain of wideband antenna and first- and second-order bandstop filtering antennas.% \red{do we need change the subcaption or the caption since both mention the simulated and measured}. \blue{I think the current format is clear/correct enough.}
}
  \label{gain}
\end{figure}

Simulated and measured E- and H-plane radiation patterns
%(at four passband and stopband frequencies) 
are depicted in Fig.~\ref{rad_UC_2_2}. As in \cite{Chen2022}, the azimuth plane measurement was limited to $\pm$130\(\degree\) due to mechanical limitations of the used hardware, which also allowed pattern measurements only up to 40\,GHz. 
%The simulated cross-polar component is not shown as it is extremely low with an ideal differential feed and a perfectly symmetric structure. 
The simulated and measured co-polar patterns agree well, but the measured cross-polar level is more noticeable. This may result from slight asymmetry related to fabrication tolerances, or be caused by feed imbalance from coaxial cable and measurement inaccuracies (e.g., AUT alignment). The radiation patterns are stable and symmetric across a wide band at both LB and UB between two principle planes due to the ME dipole operation. 
%The E- and H-plane patterns indicate a good symmetry. 
Both the simulated and measured XPD levels are high at boresight as expected due to the symmetric structure, with measured XPD of 27.4\,dB and 25.1\,dB across the frequencies of interest at LB and UB, respectively. %\red{two XPD values are for LB and UB respectively}
%Slight discrepancies between the cross-polarizations of two orthogonal ports result from fabrication tolerances and prototype alignment during the measurements.

Fig.~\ref{gain} shows that the simulated and measured realized gain of Ant-I and Ant-II agree well across respective operation bands. At LB and UB, the measured gain of Ant-II are 4.8--6.4\,dBi and 6.1--9.8\,dBi respectively. 
%while the simulated one is 4.2--6.7\,dBi.
%As for UB, the simulated gain is 6.8--8.3\,dBi whereas the measured one is 6.1--9.8\,dBi. 
Dissipation effects in the bandstop filter slightly reduce the LB gain of the proposed bandstop antenna compared to the reference antenna: a finite $Q$ (due to dielectric and metallic losses) causes some insertion loss in the resonators~\cite{hong2004microstrip}. In contrast with the measured passband peak gain, Ant-I has only 9.6\,dB of out-of-band rejection level while Ant-II gives 23.7\,dB of rejection at stopband. The upper stopband edge of Ant-II is much sharper than that of Ant-I.
%Due to hardware limitations at the time of measurement, the gain patterns were only measured up to 40\,GHz.

%Adding the two resonators sets suppresses the mid-band radiation of the reference antenna, agreeing with the discussion on radiation nulls (transmission zeros) in Section~\ref{principle}. %Radiation suppression is also implied in Fig.~\ref{Current} that the energy is stored at the resonators instead of radiation out at stopband. 
%In contrast with the measured passband peak gain, Ant-I has 9.6\,dB of out-of-band rejection level while Ant-II gives 23.7\,dB of rejection at stopband. The upper stopband edge of Ant-II element is much sharper than with Ant-I.

%\red{Some discussion should be added for Figs.~\ref{para_2_2}--\ref{para_hp}. To my understanding, they are currently not cited anywhere.} \blue{yes I will}
%As seen in Fig.~\ref{gain}, the measured UB gain is slightly higher than the simulated, which probably results from the fabrication tolerance of the air cavity size, measurement and PCB cutting accuracy and loss compensation in post processing~\cite{Chen2022}.
%The array gain is also presented in Fig.~\ref{gain} and will be discussed in Section IV-C.
%The dip in simulated gain near 28\,GHz is due to diffraction effects at the outer edge of the PCB. 
\subsection{Parametric Study}

\subsubsection{Length and location of the hairpin resonator}
\begin{figure}[!t]
\hfill
\subfigure[Resonator length.]{\includegraphics[scale=0.22]{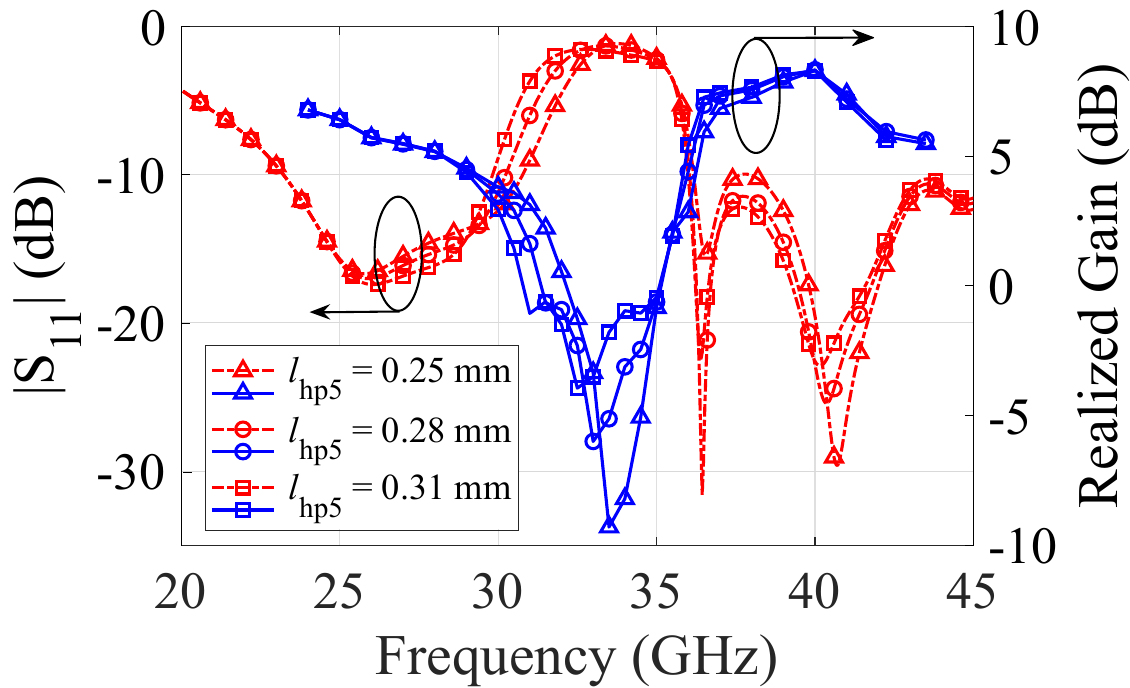}
}
\hfill
\subfigure[Resonator location.]{\includegraphics[scale=0.22]{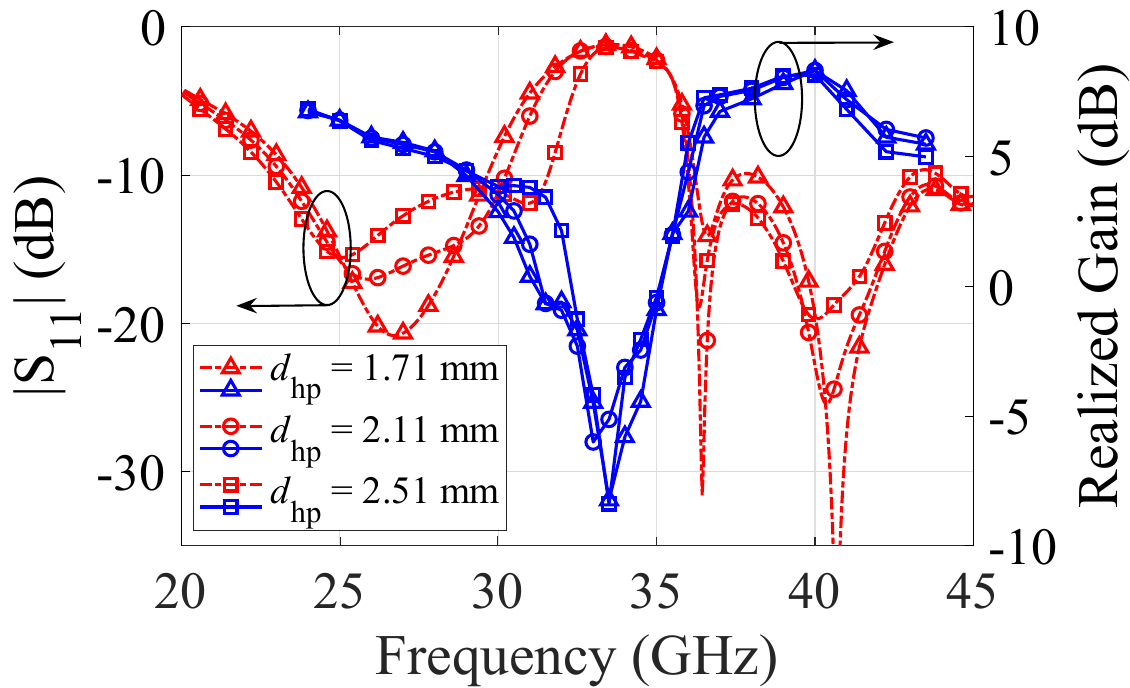}
}
\hfill
\caption{Effect of the hairpin resonator on the filtering. %\red{Remember the arrows!}
}% \red{Are these plots currently discussed anywhere?} }
\label{para_hp}
\end{figure}

Fig.~\ref{para_hp}(a)
%, the reflection coefficients and realized gain 
shows that only the lower stopband edge shifts as the hairpin resonator length changes. A longer resonator moves the first transmission zero (around 32\,GHz) generated by the hairpin resonator to lower frequency while the stopband upper bound remains fixed. The band-rejection level at stopband improves when the two transmission zeros get closer. Also, the slope of the lower stopband edge stays the same while shifting with the resonator length.

Typically, the unit elements are redundant in traditional filter synthesis. Their filtering property is not used 
%and the resultant filter is not an optimum one 
and they do not affect the filter selectivity~\cite{hong2004microstrip}.
% the electrical length of slotline affects the BW as it changes the 
In this work, the roll-off rate at stopband edges of the proposed antenna vary with the separation distance between two resonator types when modifying the location of the hairpin resonator
%relative to the patch slot 
presented in Fig.~\ref{para_hp}(b). 
%As shown in Fig.~\ref{ECM}(b), the patch slot forms a slotline whose electrical length $\theta_\mathrm{4}$ increases with larger $d_\mathrm{hp}$. 
Especially when increasing $d_\mathrm{hp}$ ($\theta_\mathrm{4}$), the lower stopband edge becomes steeper %shown in the reflection coefficient and realized gain results
and the band-rejection level remains similar. This is because adjusting the $\theta_\mathrm{4}$ changes the connecting line impedance $Z_\mathrm{04}^{\prime}$ and $Z_\mathrm{05}^{\prime\prime}$. Also the resonator location offset changes the ratio of coupled magnetic energy to the average stored energy which determines the overall $Q$ and hence the stopband BW varies. In short, the connecting transmission line
%is almost as effective as the open-circuited stub to 
impacts on the bandwidth and improve band-edge selectivity of the stopband, proving that the lines between the hairpin and coupled stub resonators are non-redundant unit cells. Thus, the proposed antenna operates as a quasi-optimum bandstop filter. 

\subsubsection{Length of the open- and short-circuited stubs}

\begin{figure}[!t]
\hfill
\subfigure[Short-circuited stub length.]{\includegraphics[scale=0.22]{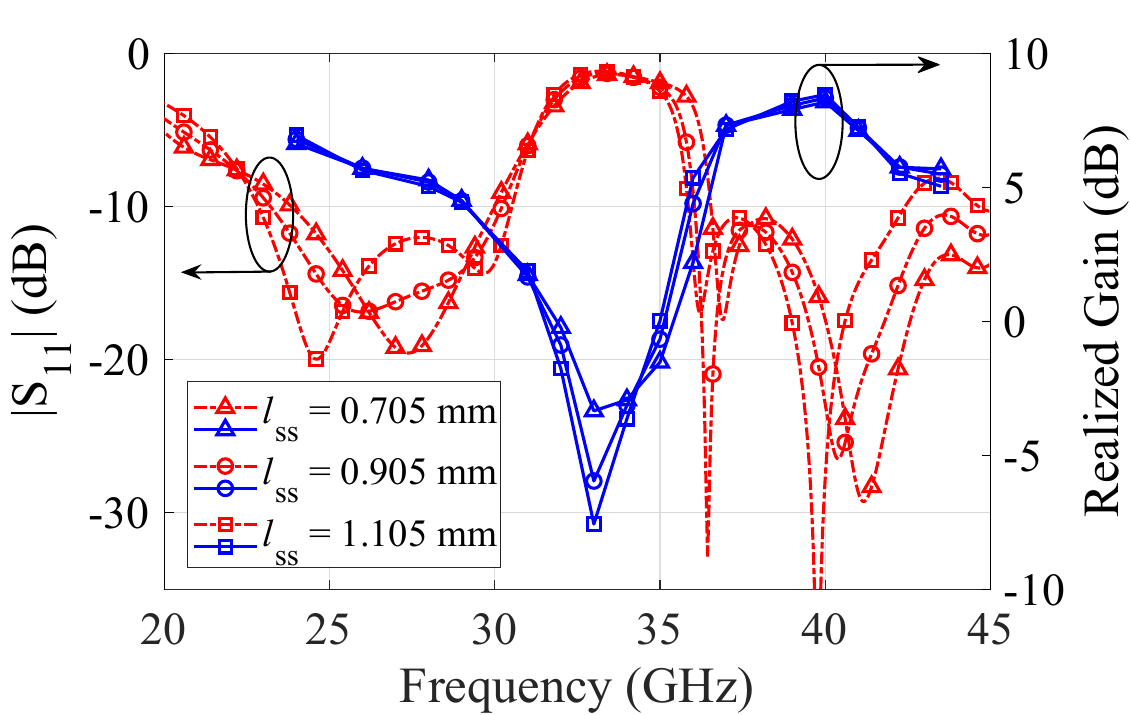}
}
\hfil
\subfigure[Open-circuited stub length.]{\includegraphics[scale=0.22]{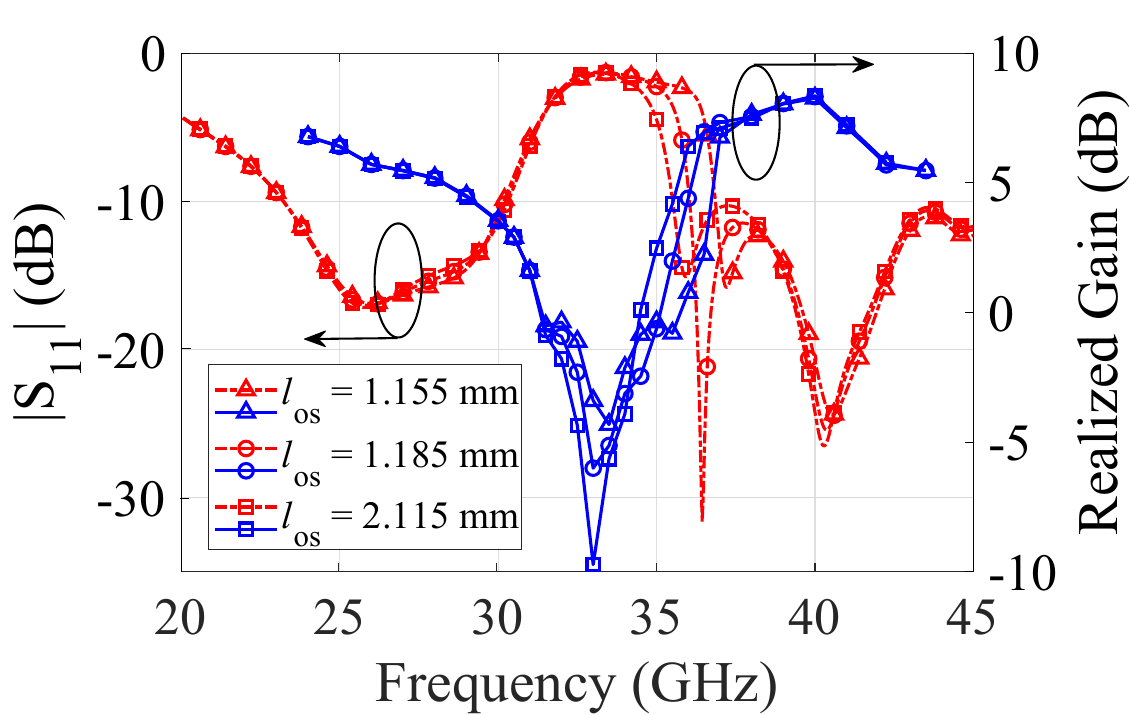}
}
\hfil
\caption{Effect of short- and open-circuited stub length on the filtering.}
\label{para_2_2}
\end{figure}

Fig.~\ref{para_2_2}(a) shows that the LB and UB matching of the proposed antenna vary with the length of the short-circuited stub $l_\mathrm{ss}$. As mentioned in Section~\ref{evo}, the transmission zero and pole locations depend on the length of open-circuited $l_\mathrm{os}$ (or $\theta_\mathrm{os}$) and short-circuited stub $l_\mathrm{ss}$ (or $\theta_\mathrm{ss}$), respectively. 
With fixed $l_\mathrm{os}$, shorter $l_\mathrm{ss}$ shifts the LB and UB transmission poles to higher frequencies at different rates with fixed transmission zeros created by the two sets of resonators. 
%In particular, the transmission poles at the left and right sides of the transmission zero detune at different rates~\cite{Chuang2012}.
%: the UB transmission pole moves faster the the lower one~\cite{Chuang2012}.
%frequency detuning of the transmission poles at the left and right side of the transmission zero are different that the location of transmission pole at UB moves faster than lower one as indicated in \cite{Chuang2012}. 
%Thus, a pair of asymmetric transmission poles can tune the LB and UB matching as shown in Fig.~\ref{para_2_2}(a) by varying $l_\mathrm{ss}$. 
%While LB has good matching, the original matching at the UB upper bound of the reference wideband and first-order band-notched antennas does not cover the desired n259 band up to 43.5\,GHz. The electrical length $\theta_\mathrm{ss}$, therefore, is tuned to primarily match at UB, e.g. moving the transmission pole on the right side of transmission zero closer to UB band edge to enhance the UB matching. 

In Fig.~\ref{para_2_2}(b), increasing the length of open-circuited stub $l_\mathrm{os}$ (or $\theta_\mathrm{os}$) shifts the upper stopband edge to lower frequency while the lower band edge remains unchanged. Hence, the length of open-circuited stub only affects the location of the second transmission zero (around 35\,GHz). In turn, the length of short-circuited stub mainly affects the location of transmission poles and the respective matching level at left and right side of the transmission zero. Varying the stub length maintains the slope of the upper stopband edge.% remains the same when shifting.

\begin{figure}[!t]
\hfill
\subfigure[Reflection coefficients.]{\includegraphics[scale=0.24]{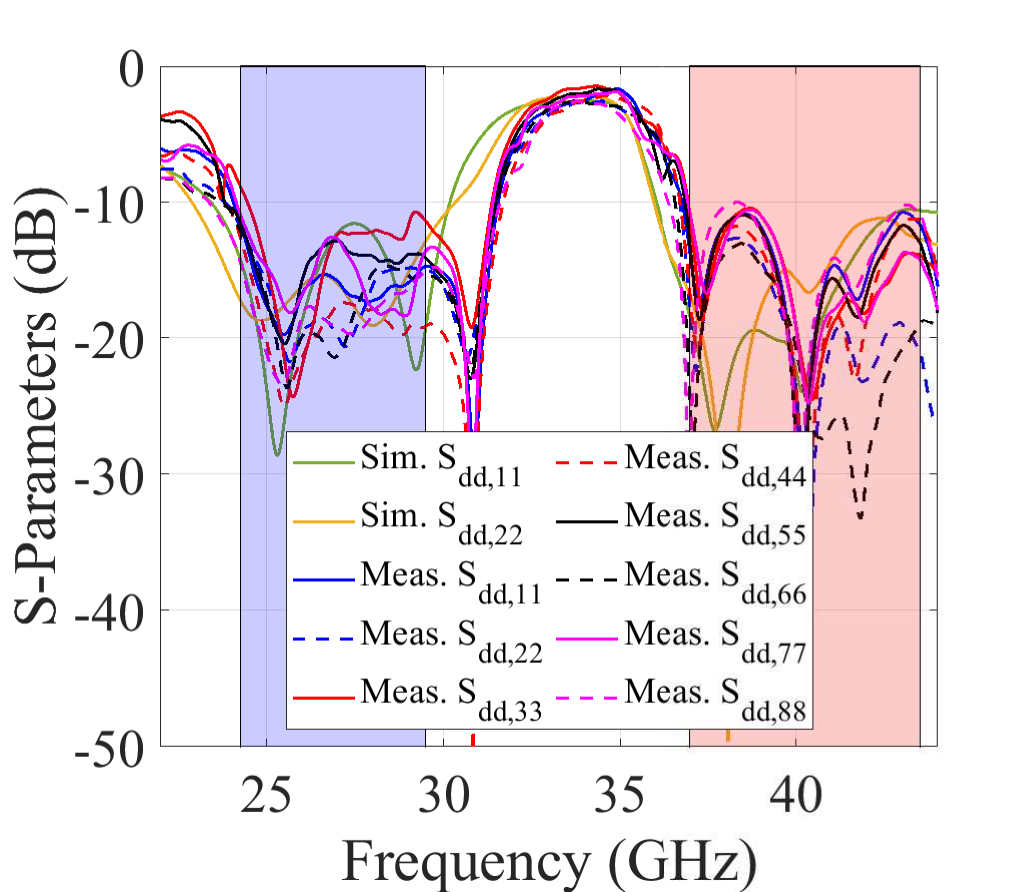}}
~
\subfigure[Coupling between adj. elements.]{\includegraphics[scale=0.24]{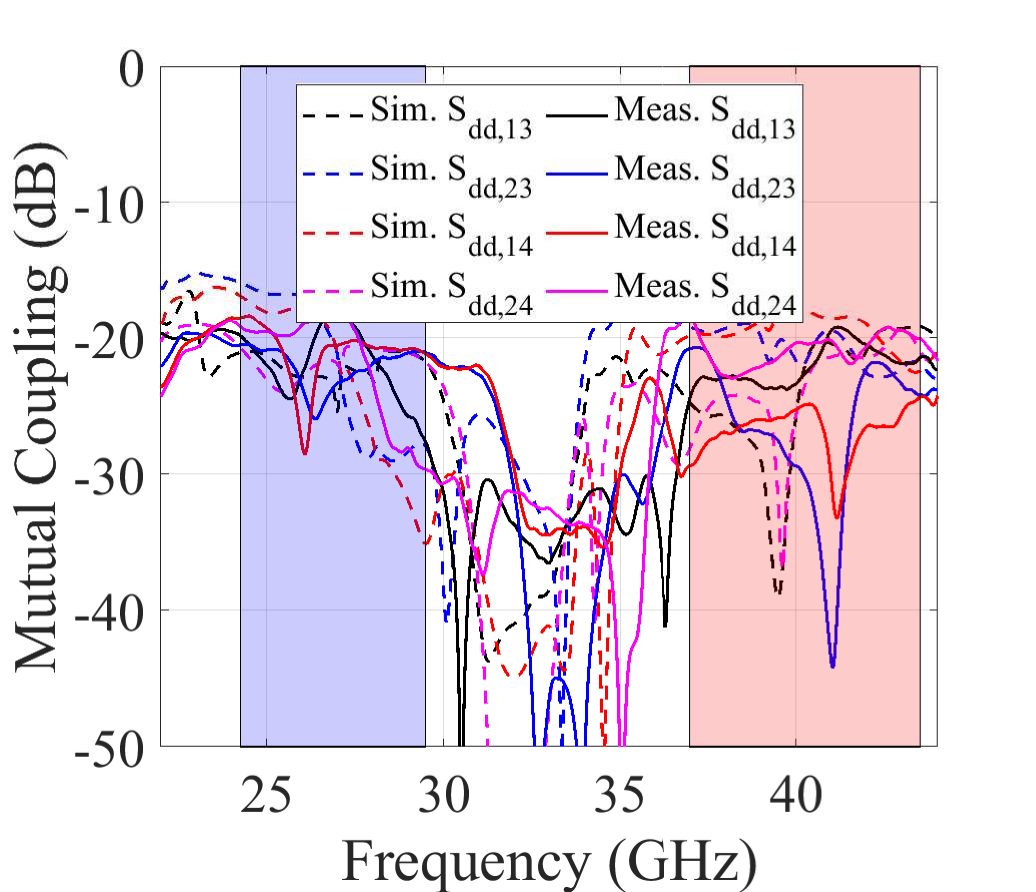}}
\hfill
%\subfigure[Coupling between diag. elements.]{\includegraphics[scale=0.243]{Plot/S_para_array_2_1_iso2_3.eps}}
\hfill
\caption{Simulated and measured $S$-parameters of the array elements.}% \red{How about just showing the worst case figure (adjacent elements) here and saving the diagonal one for the thesis?}\blue{I also think so, let me finish other parts first}} %\blue{wider figure(a)}\red{how about insert the prototype figure into this subfigure e.g. (d)?}. \blue{I think the prototype works well as a stand-alone figure. If you want to use the space more efficently, an alternative is to have, say, figure (a) as whole-column wide and below it (b)-(c) side by side. (Depending of course on whether you want to highlight matching or coupling.}}
\label{S_para_array}
\end{figure}

\subsection{2$\times$2 Array}\label{array}

Fig.~\ref{S_para_array} gives the simulated and measured $S$-parameters
%reflection coefficients 
of each unit cell antenna in the 2\(\times\)2 array. All active elements can cover the desired LB and UB for both polarizations, except for slight deterioration for differential port 3 at lower edge of LB. This effect can come from fabrication and assembly tolerances, 
%soldering uncertainties of SMPM connector
or amplitude/phase imbalances in the feed cables. %Mutual coupling between different active elements with different polarizations is given in Figs.~\ref{S_para_array}(b)-(c). 
Simulated and measured mutual coupling levels generally agree well, and adjacent elements have higher coupling than diagonal ones across the whole band. For brevity, only the worst case (isolation between adjacent elements) is shown in Fig.~\ref{S_para_array}(b). Isolation remains better than 15.9\,dB within the LB and UB. %, and it is greater for any two co-polarized elements.

\begin{figure}[!t]
\centering
\includegraphics[scale=0.50]{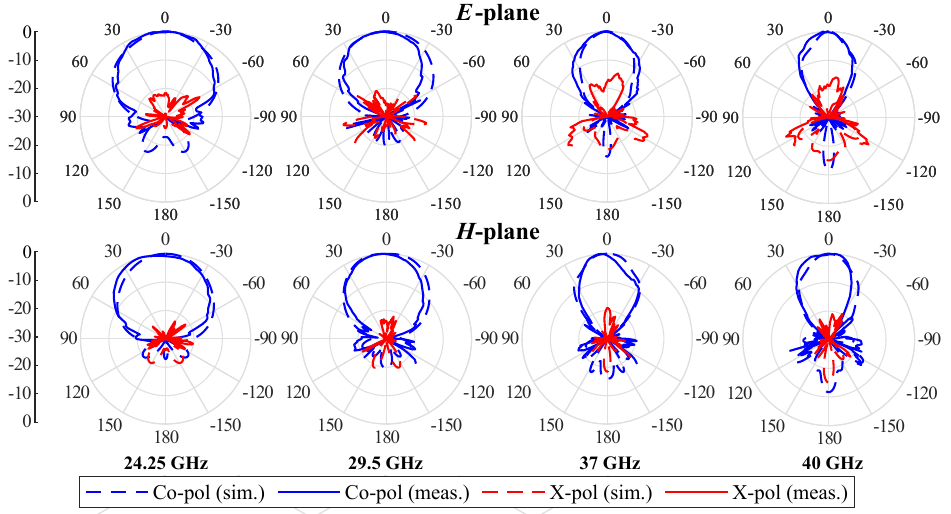}
\caption{Simulated and measured radiation patterns of the 2$\times$2 array  with second-order filtering at 24.25\,GHz, 29.5\,GHz, 37\,GHz and 40\,GHz.}%\red{how about this version (two polarizations for the array)? or do you prefer to put single element and 2x2array together?}. \blue{Looks good! I would have the single element and array in separate figures to allow more flexibility in placing them in the paper (different locations if needed etc...).}}
\label{rad_array}
\end{figure}

The simulated and measured radiation patterns for the array of Ant-II of Fig. \ref{rad_array} agree well. As the two polarizations have perfect symmetry, only one set of the patterns is provided. At E- and H-planes, the XPD level is better than 22.8\,dB at LB and 18.7\,dB at UB, which are good levels for practical applications. 
%Note that the array patterns in $xz$- and $yz$-planes are shown at diagonal planes (D-planes) instead of E- and H-planes of single element shown in Fig. \ref{Fig. 12.}.Usually, cross-polar components in D-plane are higher than those in E- and H-planes, thereby reducing the XPD in D-plane \cite{Aboserwal2018}. 
The simulated and measured array gains are illustrated in Fig.~\ref{gain}.
At LB and UB,
%the simulated gain is 8.3--10.4\,dBi while 
the measured value is 6.1--11.1\,dBi and 8.9--12.5\,dBi, respectively. 
%At UB, the simulated and measured gain values are 13.1--13.5\,dBi and 8.9--12.5\,dBi, respectively. %The mid-band radiation is suppressed after including the second-order bandstop filtering. 
%Compared with the measured passband peak gain,
The Ant-II array shows 15.5\,dB suppression level whilst the second-order one provides 21.8\,dB rejection at stopband. Gain enhancement for the array over a single element is lower at LB than at UB, probably due to stronger LB mutual coupling (see Fig.~\ref{S_para_array}). This couples more power to other (dummy and active) elements and increases the LB radiation loss \cite{Rasilainen2017, Chen2022}. Compared to simulations, the slight gain variation, beam tilt and reduced XPD in the measured patterns relate to measurement and fabrication inaccuracies. 
The measured sidelobe level at both planes is below --14.3\,dB across the whole band.% of interest.
%The measured sidelobe level at both planes is below --12\,dB across the whole band of interest.  

%For LB and UB, while the simulated array gain ranges from 8.7\,dBi to 13.9\,dBi, it ranges from 7.1\,dBi to 13.4\,dBi for the measured values. It can be seen that the single-element gain enhancement by using an array is lower at LB (from 24\,GHz to 26\,GHz) than at UB. This is probably caused by stronger mutual coupling at LB (see Fig. \ref{S_para_array}), which leads to more power coupling to other elements (both dummy and active elements) and hence larger LB radiation loss \cite{Rasilainen2017}, as well as higher cross-polarized components at LB than that at UB. The slight tilt, gain variation, and increase of the cross-polarized component level in the measured radiation patterns compared with the simulated ones can be attributed to prototype misalignment, phase and amplitude imbalance of differential feeding cables due to bending, and fabrication tolerances.  \red{How is this paragraph different from the previous one?}
%%%%%%%%%%%%%%%%%%

Table~\ref{tab2} compares the proposed dual-wideband dual-polarized antenna with the state of the art. Most reported works lack simultaneous LB and UB coverage (except \cite{Yang2022}). The proposed complementary antenna 
%with second-order bandstop filtering 
is more broadband at both LB and UB, especially at UB due to the upper transmission pole from the coupled open/short-circuited stub resonator. Hence, the proposed antenna can cover 5G bands n257, n258 and n261 (24.25--29.5\,GHz), and n259 and n260 (37--43.5\,GHz). It has a small size and much higher port-to-port isolation and XPD than most of the other reported designs. 
%Also the proposed antenna has better band-edge selectivity than that of \cite{Yang2022}.

\begin{table}[!t]
\caption{Comparison Between the Proposed Antenna and State-of-the-Art mm-Wave Antenna Designs}
\begin{center}
\begin{tabular}{c c c c c c }
\toprule

\multirow{2}{*}{\scriptsize Ref.} & \scriptsize Oper. band & \scriptsize --10-dB BW & \scriptsize Isol. & \scriptsize XPD & \scriptsize $G_\mathrm{max}$\,(dBi)\\ 

  & \scriptsize (GHz) & \scriptsize(\%) & \scriptsize(dB)  & \scriptsize(dB) & \scriptsize(UC/Array) \\ 
\hline
 
\multirow{2}{*}{\scriptsize \cite{Yang2022}}  & \scriptsize 24.0--30.0 & \scriptsize 22.2\% & \scriptsize $>$20  & \scriptsize $>$20 & \scriptsize  --/7.1 \\ 

  & \scriptsize 37.0--43.5 & \scriptsize 16.1\% & \scriptsize $>$20   & \scriptsize $>$20 & \scriptsize --/8.2 \\ 
\hline

\multirow{2}{*}{\scriptsize \cite{Li2018}} & \scriptsize 23.7--29.2 & \scriptsize 20.7\% & \multirow{2}{*}{\scriptsize N/A}  & \scriptsize $>$25\,(sim.) & \scriptsize 7.2/-- \\ 

 &  \scriptsize 36.7--41.1 & \scriptsize 11.3\% & \scriptsize & \scriptsize $>$25\,(sim.) &  \scriptsize 10.9/--  \\ 
\hline
 %\midrule

%%%%%%%%%%%%%%%%%%

\multirow{2}{*}{\scriptsize \cite{Deckmyn2019SIW}} & \scriptsize 26.7--30.4 & \scriptsize 12.8\% & \multirow{2}{*}{\scriptsize N/A} & \scriptsize \multirow{2}{*}{\scriptsize N/A} &\scriptsize --/10.1   \\ 

 & \scriptsize 36.6--38.8 & \scriptsize 5.8\% & \scriptsize  &   &  \scriptsize--/10.2 \\ 
\hline
 %\midrule
%%%%%%%%%%%%%%%%%%%

\multirow{2}{*}{\scriptsize \cite{FengMeta2020}} & \scriptsize 23.8--27.7 & \scriptsize 15.2\% & \scriptsize 35  & \scriptsize 20.1 & \scriptsize 7.7/11.0 \\ 

 & \scriptsize 36.8--44.4 & \scriptsize 18.7\% & \scriptsize 28 & \scriptsize24.7 & \scriptsize 13.0/15.5  \\ 
\hline
 %\midrule
%%%%%%%%%%%%%%%%%%%

\multirow{2}{*}{\scriptsize \cite{Sun2021Dualband}} & \scriptsize 23.3--31.7 & \scriptsize 30.5\% & \scriptsize 17.1 & \multirow{2}{*} {\scriptsize N/A} & \scriptsize --/14.8 \\ 

   & \scriptsize 42.5--46.5 & \scriptsize 9.0\% & \scriptsize 18.3 &  & \scriptsize--/14.1 \\ 

\hline
 %\midrule
%%%%%%%%%%%%%%%%%%%
\multirow{2}{*}{\scriptsize \cite{Zeeshan2023}}  & \scriptsize 24.25--29.5 & \scriptsize 19.5\% & \scriptsize 20 & \scriptsize 20 & \scriptsize 3.5/7.0 \\ 

 & \scriptsize 37.0--40.0 & \scriptsize 7.8\% & \scriptsize 20   &\scriptsize 20 & \scriptsize 4.5/10.0  \\ 
\hline
% \midrule
%%%%%%%%%%%%%%%%%%%

\multirow{2}{*}{\textbf{\scriptsize This work}}  & \scriptsize 23.9--31.6 & \scriptsize 27.7\% & \scriptsize 36.4 & \scriptsize 27.4 & \scriptsize  6.7/11.1 \\ 

  & \scriptsize 36.7--45.0 & \scriptsize $>$20.3\% & \scriptsize 31.6    & \scriptsize 25.1 & \scriptsize 8.3/12.5  \\

\bottomrule
%\multicolumn{6}{l}{\scriptsize {$^{(a)}$Matching criterion: --10\,dB \hskip 3mm $^{(b)}$Unit cell / Array} \hskip 3mm %UC\,=\,Unit Cell
%}\\
\end{tabular}
\label{tab2}
\end{center}
\end{table}

%%%%%%%%%%%%%%%%%%
\section{Conclusion}
A dual-wideband differentially-fed dual-polarized magnetoelectric (ME) dipole with second-order bandstop filtering operation for Ka-band applications is presented. The proposed antenna covers 5G bands n257, n258 and n261 (24.25--29.5\,GHz), and n259 and n260 (37--43.5\,GHz). Independently controlled transmission zeros and poles give sharp selectivity and allow tuning the stopband BW and rejection level. The design keeps a compact footprint and has high port-to-port isolation of 36.4\,dB and 31.6\,dB and XPD of 27.4\,dB and 25.1\,dB at LB and UB, respectively. The rejection level for the proposed single element is as high as 23.7\,dB. With the favorable out-of-band rejection, compact size, simple structure and low-cost fabrication, the proposed ME dipole is a promising candidate for mm-Wave Antenna-in-Package (AiP) applications. 
\section*{Acknowledgment}
% The authors would like to thank Aspocomp Group for the PCB manufacturing, and Mr. Markku Jokinen for help in the radiation pattern measurements. 

The authors would like to thank Mr. Markku Jokinen for help in the radiation pattern measurements. Keysight Inc. has supported the research by donating measurement equipment.

%The authors thank Aspocomp Group for PCB manufacturing and M. Jokinen for help in the radiation pattern measurements. Keysight Inc. has donated measurement equipment.

\ifCLASSOPTIONcaptionsoff
  \newpage
\fi

\bibliographystyle{IEEEtran}
\bibliography{IEEEabrv,bibliography}
\end{document}